# Low-Dose 3D Bonding Mapping Through "Soft" Core-Loss EELS Tomography and Unsupervised Deep Learning

Mario Pelaez-Fernandez[1,2*], Daniel del-Pozo-Bueno[3], Adrien Teurtrie[1,4], Serge Brosset[3], Maya Marinova[5], Phillipe Ciuciu[6,7], Marta Estrader[8,9], German Salazar-Alvarez[10], Francesca Peiró[9,11], Raul Arenal[2,12], Sonia Estradé[9,11], Zineb Saghi[3], Francisco De la Peña[1*]

[1] Unité Matériaux et Transformations (UMET UMR 8207), Université de Lille, Bâtiment C6, Cité Scientifique, Villeneuve d'Ascq, 59650, France

[2] Instituto de Nanociencia y Materiales de Aragon (INMA), CSIC-U. de Zaragoza, Calle Pedro Cerbuna 12, Zaragoza, 50009, Spain

[3] Univ. Grenoble Alpes, CEA, Leti, F-38000 Grenoble, France

[4] Centre d'Elaboration de Matériaux et d'Etudes Structurales (CEMES), University of Toulouse and CNRS, 31055 Toulouse, France

[5] Univ. Lille, FR 2638-IMEC-Institut Michel-Eugène Chevreul, F-59000 Lille, France

[6] CEA, Joliot, NeuroSpin, Gif-sur-Yvette Cedex 91191, France.

[7] Inria, MIND, Université Paris-Saclay, Palaiseau 91120, France.

[8] Department of Inorganic and Organic Chemistry, Inorganic Chemistry Section, University of Barcelona, Carrer de Martí i Franquès, 1-11, 08028 Barcelona, Spain

[9] Institute of Nanoscience and Nanotechnology (IN2UB), Universitat de Barcelona, 08028 Barcelona, Spain.

[10] Department of Materials Science and Engineering, Ångström Laboratory, Uppsala University, Uppsala 751 03, Sweden

[11] LENS-MIND, Departament d'Enginyeria Electrònica i Biomèdica, Universitat de Barcelona, 08028, Barcelona, Spain

[12] Laboratorio de Microscopias Avanzadas, Universidad de Zaragoza, Calle Mariano Esquillor, Zaragoza, 50018, Spain

**emails: mariopf@unizar.es, francisco.de-la-pena-manchon@univ-lille.fr*

## Abstract

Resolving the three-dimensional (3D) chemical configuration of beam-sensitive nanomaterials at high spatial resolution remains a persistent frontier in Scanning Transmission Electron Microscope (STEM). The main limitation usually lies in the trade-off between the high electron dose required to obtain meaningful analytical signals and the large number of projections needed for reliable tomographic reconstruction. Here, we overcome these constraints to achieve dose-efficient 3D bonding mapping of beam-sensitive FeO/$Fe_3O_4$ core-shell nanocubes with unprecedented resolution via electron energy loss spectroscopy (EELS) analyses.

Our approach relies on two key technical developments. First, we develop a standardless “soft” core-loss EELS methodology that exploits the high-cross-section shallow ionization edges (Fe-$M_{2,3}$), delivering ~50x greater dose efficiency than conventional Fe-$L_{2,3}$-edges, using the latter only as a reliable source to obtain standards for FeO and $Fe_3O_4$. Second, we introduce an unsupervised deep learning method -multi-channel deep image prior with total variation regularization (DIPm-TV) for spectroscopic tomography reconstruction. The method demonstrates superior performance under highly sparse and low-dose acquisition conditions. Using simulated datasets, we show that high-quality reconstructions can be obtained from as few as nine low-dose projections acquired over a tilt range of -70° to +70°, without relying on the high-angle annular dark-field (HAADF) STEM signal or imposing symmetry constraints.

The exploitation of the Fe-$M_{2,3}$ edges on FeO/$Fe_3O_4$ nanocubes has resulted in EELS maps with a much better signal-to-noise ratio and resolution compared to state-of-the-art, revealing fine structural features, including a thin outer shell of FeO surrounding the magnetite shell. The 3D reconstruction of the soft EELS tilt series using DIPm-TV yields high-quality oxidation-state reconstructions, with an isotropic resolution of approximately 1 nm. The reconstructed volumes preserve the cubic morphology of the FeO/$Fe_3O_4$ core-shell nanocubes, successfully recover the thin outer layer, and reveal a small internal void within the shell structure. While some of these features are faintly visible in individual 2D projections, they remain inaccessible using conventional reconstruction methodologies.

Overall, this work establishes a new pathway for low-dose 2D and 3D analytical mapping of beam-sensitive materials containing elements with shallow core-loss (CL) ionization edges, enabling orders of magnitude dose reduction while maintaining spectral fidelity and reliable 3D information.

## Introduction

Over the past two decades, the scanning transmission electron microscope (STEM) has emerged as one of the premier tools for materials science characterization, offering the unique ability to capture multi-modal signals (encompassing classical imaging, spectroscopy and 4D-STEM) at resolutions approaching tens of picometers (Pennycook & Nellist, 2011). Historically, progress in STEM was driven primarily by hardware innovations, such as improved electron sources, aberration correctors, and monochromators. More recently, computational developments have played an increasingly central role, with breakthroughs in spatial resolution now frequently driven as much by algorithms as by hardware improvements (Nguyen et al., 2024; Qiu et al., 2025). This is in part due to the fact that, for a wide range of samples of interest, the limit is not the nominal resolution of the instrument, but the amount of information on the sample it can gather before beam damage occurs (Chen et al., 2020; Egerton, 2013, 2019; Lyu et al., 2023).

STEM electron energy-loss spectroscopy (EELS) stands out for its unique combination of high spatial and energy resolution, achieving atomic-scale resolution and energy resolutions of up to 5 meV with the latest aberration correctors and monochromators (Krivanek et al., 2009, 2015; Pattison et al., 2025). For chemical analysis, its superior energy resolution compared to energy-dispersive X-ray (EDX) spectroscopy offers insight into the material's electronic structure through its energy loss near-edge structure (ELNES) features (Arenal et al., 2008). Indeed, ELNES analysis provides information beyond elemental quantification, including oxidation states (e.g., $Fe^{2+}$ vs. $Fe^{3+}$), coordination environments (tetrahedral vs. octahedral), and bonding character (e.g., oxygen functional groups) (Bernier et al., 2008; Derikvandi et al., 2023; Lin et al., 2014; Pelaez-Fernandez et al., 2021; Poizot et al., 2000), getting closer to a "synchrotron in the microscope" with each new development (Ramasse, 2017). The applications of this characterization technique are diverse, ranging from low-dimensional materials to catalysts and battery materials (Bernier et al., 2008; Derikvandi et al., 2023; Lin et al., 2014; Pelaez-Fernandez et al., 2021; Poizot et al., 2000). Among these, one of the most prominent uses has been the determination of oxidation states in oxides (Brown et al., 2017; del-Pozo-Bueno et al., 2021; Gloter et al., 2017; Kurata & Colliex, 1993; Makarchuk et al., 2025; Torruella et al., 2016; van Aken & Woodland, 1999). For such applications, EELS remains the only technique capable of providing the necessary spatial and energy resolution, highlighting the importance of developing new approaches to further expand the range of materials it can be used on.

However, the core-loss (CL)-EELS cross-section is typically orders of magnitude lower than the elastic cross section used for 4D-STEM analysis. Therefore, beam damage severely limits the application of EELS to beam sensitive materials, such as metal-organic frameworks (MOFs), covalent-organic frameworks (COFs), and 2D nanomaterials (Chen et al., 2020; Xu et al., 2025). This is particularly pronounced in ELNES experiments: they require slightly higher electron doses for analysis, and beam-induced damage alters the sample's chemistry, consequently modifying the ELNES features being analyzed (Davoisne et al.,

2017).

In recent years, significant progress has been made in reducing the dose required for analysis and minimizing electron beam damage, through advances in both instrumentation and data processing. On the instrumentation side, several developments have contributed to this progress in STEM-EELS. Changes in acquisition conditions and electron optics, such as lowering the electron beam energy, have proven to lower knock-on damage for sensitive materials (Dumaresq et al., 2024). Direct electron detectors (DEDs) have improved detection efficiency and decreased readout time, to the point where the speed-limiting factor becomes the scanning coils of the microscope itself, enabling analysis at much lower doses (Auad et al., 2022; Faruqi & McMullan, 2018; Jannis et al., 2021). Monochromators have sharpened spectral features in EELS, allowing reliable detection at reduced dose levels (Kimoto, 2014; Krivanek et al., 2019; Lopatin et al., 2018). Cryogenic cooling has proven effective in mitigating radiolysis, while aloof beam excitation has enabled vibrational EELS measurements without direct irradiation of the specimen (Crozier, 2017). Non-uniform scanning strategies, such as Lissajous and random scan patterns, particularly when combined with event-driven detectors, have also shown promise in reducing beam damage by distributing the dose more evenly and allowing time for partial recovery of the material between successive exposures (Auad et al., 2026).

In parallel, advances in data processing have enabled analysis from dose-limited datasets. Low-rank matrix estimation methods, such as singular value thresholding (SVT) (often called PCA-principal component analysis- in the field), enables the analysis of low-signal datasets by leveraging redundancy across dimensions (Blanco-Portals et al., 2022; de la Peña et al., 2011). When combined with blind source separation (BSS), it can perform unsupervised ELNES analysis, even in cases that are challenging for conventional methods (Torruella et al., 2016). More recently, deep learning (DL) approaches have demonstrated remarkable performance in denoising, inpainting, data augmentation and super-resolution tasks, further pushing the boundaries of what can be achieved under strict dose constraints (Browning et al., 2023; del-Pozo-Bueno et al., 2024; Monier et al., 2020).

Dose effectiveness is an even greater issue when extending EELS analysis to 3D, which is of great interest for samples in which the interpretation of 2D results is hampered by projection effects. EELS tomography gives access to the 3D chemical and electronic information from a series of EELS spectral images acquired at different tilt angles. However, the amount of signal required for full tilt series acquisition substantially increases sample damage (Torruella et al., 2016), highlighting the need for sophisticated reconstruction methods specifically designed to reduce the number of required projections (sparse-view) and to perform well under low-dose acquisitions, where measurements are inherently affected by strong Poisson noise and low SNR.

Most reconstruction methods have relied on compressed-sensing (CS)-based approaches, exploiting sparsity in a transformed domain to enable reconstruction from undersampled datasets. However, their performance degrades under highly sparse and low-dose acquisition conditions (Jiang et al., 2018), requiring, in most of these works, strong prior assumptions, such as symmetry constraints. Nicoletti et al. (Nicoletti et al., 2013) successfully reconstructed surface plasmon modes in silver nanocubes using CS-based wavelet sparsity and symmetry constraints. Similarly, Li et al. (X. Li et al., 2021) achieved 3D vectorial imaging of surface phonon polaritons, using an eigenmode-based reconstruction algorithm assuming cubic symmetry, and Torruella et al. (Torruella et al., 2016) combined CL-EELS, blind source separation (BSS) and CS-based reconstruction, also under symmetry assumptions, to map in 3D the distribution of $Fe^{2+}$/$Fe^{3+}$ oxidation states in FeO/$Fe_3O_4$ core-shell nanocubes. Notably, this work investigated the same nanocube samples studied in the present work.

For more complex structures, where such assumptions are not valid, or for beam-sensitive materials, for which even fewer acquisitions are possible, multimodal (MM) reconstruction strategies have been proposed. These combine spectroscopic signals (EELS/EDX) with HAADF-STEM tilt series, using HAADF as a high-SNR structural prior to improve reconstruction quality and mitigate sparse-view and low-dose limitations (Huber et al., 2019; Kormilina et al., 2026; Manassa et al., 2025; Schwartz et al., 2024; Zhong et al., 2017, 2018).

More recently, DL strategies have been increasingly explored for low-dose spectroscopic STEM tomography, building upon similar MM frameworks. Among the most relevant works for the present study, Cha et al. (Cha et al., 2022) addressed sparse-view, low-dose STEM-EDX tomography of supported nanocrystals by acquiring simultaneous HAADF-STEM and EDX tilt series from −60° to +60° with 10° tilt increments. Their method used a CycleGAN to denoise chemical maps by leveraging correlations between EDX and HAADF signals, and a ProjectionGAN to reduce reconstruction artefacts. Overall, these approaches rely on HAADF as a high-SNR structural prior. However, in many samples, including ours, HAADF contrast is insufficient for this purpose.

In this study, we instead apply a deep image prior (DIP) approach, a training-free strategy in which an untrained convolutional neural network acts as an implicit regularizer, optimized directly on the acquired projections to enable simultaneous 3D reconstruction and restoration (Baguer et al., 2020; Lempitsky et al., 2018). This strategy was previously applied in ET by Brosset et al. (Brosset et al., 2026) for STEM-HAADF data, to reconstruct the 3D morphology of Pt nanoparticles under limited-angle and sparse-view conditions, showing competitive performance with supervised methods when combined with total variation (TV) regularization, while highlighting DIP's advantage of requiring no training data.

In this work, we extend this DIP-based methodology to low-dose, highly sparse-view STEM-EELS tomography, introducing a multi-channel approach that leverages the correlation between the nanoparticle core and shell EELS signals to enhance the quality of the 3D oxidation state mapping. An important advancement of this approach is that it does not require either HAADF tilt series nor external training data. The reconstruction is driven solely by the EELS measurements through the forward model, with regularization provided by the DIP architecture and the explicit TV term, which confer robustness against noise and severe angular undersampling. We validate DIPm-TV on a simulated core-shell structure using just nine projections over a -70° to +70° tilt range, achieving high-fidelity reconstructions while suppressing artifacts associated with large tilt increments, a mild missing wedge and noisy acquisitions. We then apply the DIPm-TV framework to experimental Fe-$M_{2,3}$ EELS intensity maps.

In addition to this novel reconstruction strategy, we also address dose limitations at the acquisition level by introducing a dose-efficient spectroscopic approach. In this context, this study introduces oxidation state mapping using ELNES analysis on low-energy CL EELS edges, which offer much higher cross-sections than the commonly used medium- and high-energy edges, optimizing the SNR per electron impact. This approach, which we have named Soft Core-Loss EELS (or Soft EELS for short), enables EELS and ELNES characterization at a fraction of the electron dose. Previous studies have employed these lower energy edges to differentiate compounds containing the same element (Van Aken & Liebscher, 2002; van Aken & Woodland, 1999) and for elemental mapping of elements lacking CL edges at higher energies, such as lithium (Park et al., 2021). However, no ELNES-based mapping in this energy range has been reported to date, to the best of our knowledge.

The complexity of acquiring, analyzing and mapping CL edges at low energy has hindered the application of Soft EELS. Indeed, these edges lie at the tail of the volume plasmon, which is typically one of the most intense features of the EELS spectrum after the zero-loss peak. For reliable analysis in this spectral range without multiple exposures, the whole low-loss (LL) spectrum must be acquired with sufficient dynamic range to enable a good SNR of the CL features of interest, without saturating the zero-loss peak. This has only been possible recently with the integration of the new generation of hybrid pixel direct detectors in EELS spectrometers. Moreover, EELS analysis in this energy range is notoriously difficult. Indeed, the proximity of the volume plasmon makes the standard power-law background model unreliable, while the non-linearity of the EELS signal in this energy range precludes the usage of BSS methods, which rely on linearity.

Combining these two approaches, this study aims to provide a proof-of-concept for the application of soft EELS tomography, using the Fe-$M_{2,3}$ edges at 54 eV, instead of the widely used Fe-$L_{2,3}$ at 708 eV, as an example. The experiment replicates a previous ELNES tomography study on core-shell $FeO/Fe_3O_4$ nanocubes (Torruella et al., 2016), but using state-of-the-art acquisition and analytical techniques to push the limits of this characterization. The original study faced two major challenges. First, beam damage severely limited the size of the EELS tilt-series, requiring symmetry constraints for tomographic reconstruction, with only half of the tilt-series not being affected by beam damage. Second, the impossibility of acquiring the zero loss peak and the CL-EELS spectra on that particular experiment required an approximation to account for multiple scattering (Torruella et al., 2016, Section Supplementary Information) that cannot be easily applied to more complex samples.

Soft EELS analysis and DIPm-TV tomography provide solutions to both these issues. Using the Fe-$M_{2,3}$ significantly improves the SNR, while requiring a single spectrum for analysis. DIPm-TV enables the reconstruction under low-dose and highly sparse acquisition conditions, without compromising spatial information, enabling high-quality, high-SNR reconstructions with isotropic resolution, representing a significant advance over state-of-the-art methods while further reducing the required electron dose. This opens a new and exciting possibility for the 3D study of chemical properties at the nanoscale, particularly for beam-sensitive materials with ionization edges in this energy range.

## *Materials and Methods*

### Materials

The $FeO@Fe_3O_4$ core-shell nanocubes are from the same batch of those in (Estrader et al., 2015). As shown in previous studies (Torruella et al., 2016), the $Fe_3O_4$ shell of the nanocube has grown epitaxially on the FeO core. FeO and $Fe_3O_4$ yield nearly identical HAADF signals, as their densities differ by only ~10%, highlighting the need for ELNES analysis (see Figure 1).

These samples were stored in air for several years and present greater challenges than the pristine specimens from the original study, exhibiting degradation features such as core shrinkage. Prior to STEM-EELS analysis, the samples were thermally treated overnight at 70°C under vacuum., to prevent possible organic pollution during the STEM-EELS experiments.

## STEM-EELS acquisition

STEM-EELS measurements were taken using a Cs aberration-corrected and monochromated Thermo Fisher Scientific Titan Themis (S)TEM. The microscope was equipped with a Gatan Quantum 966 ERS GIF equipped with a Quantum Detectors MerlinEELS camera, consisting of 4 MEDIPIX3 hybrid-pixel DEDs, for a total 1024x256 pixels, and a 500 µm silicon sensor. The MerlinEELS is operated through a NionSwift plugin. A custom NionSwift Python plugin implements the acquisition of quasi-simultaneous multi-energy and/or multi-frame spectrum images (SPIMs).

Three distinct EELS datasets have been acquired for these experiments, one comprised of two quasi-simultaneous LL and CL SPIMs for extracting the Fe-$L_{2,3}$ and Fe-$M_{2,3}$ fingerprints (hereafter called F-CL and F-LL), one LL SPIM for 2D bonding mapping using the Fe-$M_{2,3}$ fingerprints (which we will call M-LL), and a LL tilt-series comprised of 9 SPIMs (henceforth T-LL) for 3D bonding mapping. All datasets were acquired at 200 keV, with a convergence angle of 17 mrad, a collection angle of 49 mrad, and a 70keV threshold on the MerlinEELS camera. The full-width half-maximum of the energy spread of the electron was 0.4 eV. We also acquired a gain reference of the camera at the same beam energy and threshold. The LL spectral range was -16:187 eV for all datasets. The corresponding HAADF images were acquired for each SPIM acquisition.

The F-LL/F-CL dataset was acquired in multi-energy and multi-frame mode to enable the quasi-simultaneous acquisition of the LL and CL spectra while minimizing and monitoring beam damage. In particular, a series of 5 alternating LL and CL SPIMs were acquired on a 111x188 grid with a dwell time of 5 ms and an energy dispersion of 0.2 eV, resulting in 2 datasets of 111x188x5x1024 pixels each. Their spectral ranges were [-16,187] eV and [684,887] eV respectively. The pixel size was 0.5 nm. The dose rate for these acquisitions was 0.20 $pA \cdot Å^{-2} \cdot s^{-1}$, with a total dose of $\sim 6 \cdot 10^4$ $e^- \cdot Å^{-2}$.

The M-LL 132x123x1024 pixels SPIM was acquired in a single frame on a different area of the sample. This SPIM was acquired using an energy dispersion of 0.1 eV in the -12:90 eV energy range, with a dwell time of 1 ms and a pixel size of 0.3 nm. The dose rate for these acquisitions was 4.22 $pA \cdot Å^{-2} \cdot s^{-1}$, with a total dose of $\sim 2.3 \cdot 10^4$ $e^- \cdot Å^{-2}$.

The T-LL tilt-series dataset was acquired in single-frame mode, and consists of nine 125x125x1024 pixels SPIMs acquired at -69.5º, -52.5º, -35º, -17.5º, 0º, 17.5º, 35º, 52.5º, and 70º tilt-angles. The pixel size was 0.4 nm and the dwell time 1ms. The dose rate for the tomographic studies was 0.44 $pA \cdot Å^{-2} \cdot s^{-1}$, with a total dose of ~4100 $e^- \cdot Å^{-2}$ for each of the 9 acquisitions for a total cumulated dose of $\sim 3.1 \cdot 10^4$ $e^- \cdot Å^{-2}$.

## STEM-EELS analysis

Our main objective is to perform EELS bonding mapping with the Fe-$M_{2,3}$ edges rather than the Fe-$L_{2,3}$ edges to solve the core-shell structure of the iron oxide nanocubes that we have investigated while minimizing sample damage. EELS bonding mapping with the Fe-$L_{2,3}$ edges is usually performed either with the standards-based fingerprinting method, or BSS. In this case, the use of standards is delicate, since FeO is non-stoichiometric and unstable at room temperature in its pure form. Furthermore, external references can be sensitive to experimental conditions, such as sample thickness and experimental acquisition parameters. Consequently, a standardless BSS-based method was chosen for this analysis, following the approach of previous studies (De La Peña et al., 2011; Torruella et al., 2016). This method offers advantages of simplicity and accuracy, since it does not require spectral reference and calibrations for analysis.

However, after initial acquisitions, it was seen that conventional standardless analysis could not be used in Soft EELS the same way it is commonly used in higher energies datasets (See Supporting information), due to the complexity of the EELS signal at lower energy losses. This prompted the development of an alternative method that enables the extraction of Fe-$M_{2,3}$ fingerprints from a pair of LL and CL SPIMs from exactly the same area of the sample under study. We later used these fingerprints for the analysis of the M-LL and T-LL datasets.

EELS data were analyzed using Python scripts, as well as the HyperSpy/eXSpy software package (Francisco de la Peña et al., 2026). All SPIMs were gain-corrected using a gain-reference acquired by homogeneously illuminating the detector with the same beam energy and camera settings as the experiments. The offset of the spectra was adjusted using the position of the zero-loss peak. In the case of F-LL/F-CL datasets, a 4D datacube (*frame, x, y,* $E_{Loss}$) was created for each LL/CL pair by combining all their respective multi-frame acquisitions. Each LL/CL pair was spatially aligned using the corresponding HAADF micrographs, to account for spatial drift.

We start by performing BSS using a combination of singular value decomposition (SVD) (weighted for Poisson noise) (Keenan & Kotula, 2004) and independent component analysis (ICA) (de la Peña et al., 2011; Torruella et al., 2016) on the F-CL dataset. Because of the low number of counts of the dataset, we perform an 8 by 8 spatial binning to optimize the accuracy of the weighted SVD for Poisson noise. Next we perform weighted SVD, masking everything but the particle of interest to optimize the performance of the algorithm. Through the inspection of the scree plot of the SVD decomposition we determine that the dimension of the F-CL dataset is 3 (see Figure S2 in the supporting information). Next, we perform BSS using the FastICA algorithm (Hyvärinen & Oja, 2000) with 3 spectral components with their corresponding maps (see Figure 2): the background to the Fe-$L_{2,3}$ edge, and two distinct Fe-$L_{2,3}$ edges, that correspond to the signatures of the core and the shell of the NPs. Based on previous studies on these nanoparticles, as well as similar nanomaterials, these two signatures correspond, respectively, to the ELNES fingerprints of FeO and $Fe_3O_4$ (Brown et al., 2017; Estrader et al., 2015; Golla-Schindler et al., 2003; Torruella et al., 2016; Van Aken et al., 1998; Vigliaturo et al., 2019). These components from the BSS decomposition, once normalized using their Hartree-Slater cross-sections, directly yield the proportions of the two Fe species in the SPIM. The resulting spectra, after normalization and power-law background removal to account for the tail of the O-K and Fe-M edges at lower energies (see Figure 2), were taken as the reference Fe-$L_{2,3}$ fingerprints for the core ($L_{FeO}$) and the shell ($L_{Fe3O4}$) of the NPs.

The core principle underlying our method is that the relative abundance of the iron species (the core and the shell) derived from the normalized, background-subtracted core-loss EELS and Soft EELS edges must be strictly identical at any given spatial location. It is important to note that, even though this study only needs to separate two different species, this method is, in principle, generalizable to any number of components, as long as the dataset features enough non-correlated data points.

To achieve reliable fingerprint extraction and stabilize the background fitting, the F-CL and F-LL datasets, binned as described above, first underwent a noise reduction step that consists of applying weighted singular value thresholding (SVT) to both datasets. The signal comparison between raw, binned and SVT-denoised spectra, taking 10 and 3 components respectively, can be seen in Figure 1. For our low-count datasets, this denoising step is crucial for accurately fitting background models, particularly over narrow spectral windows.

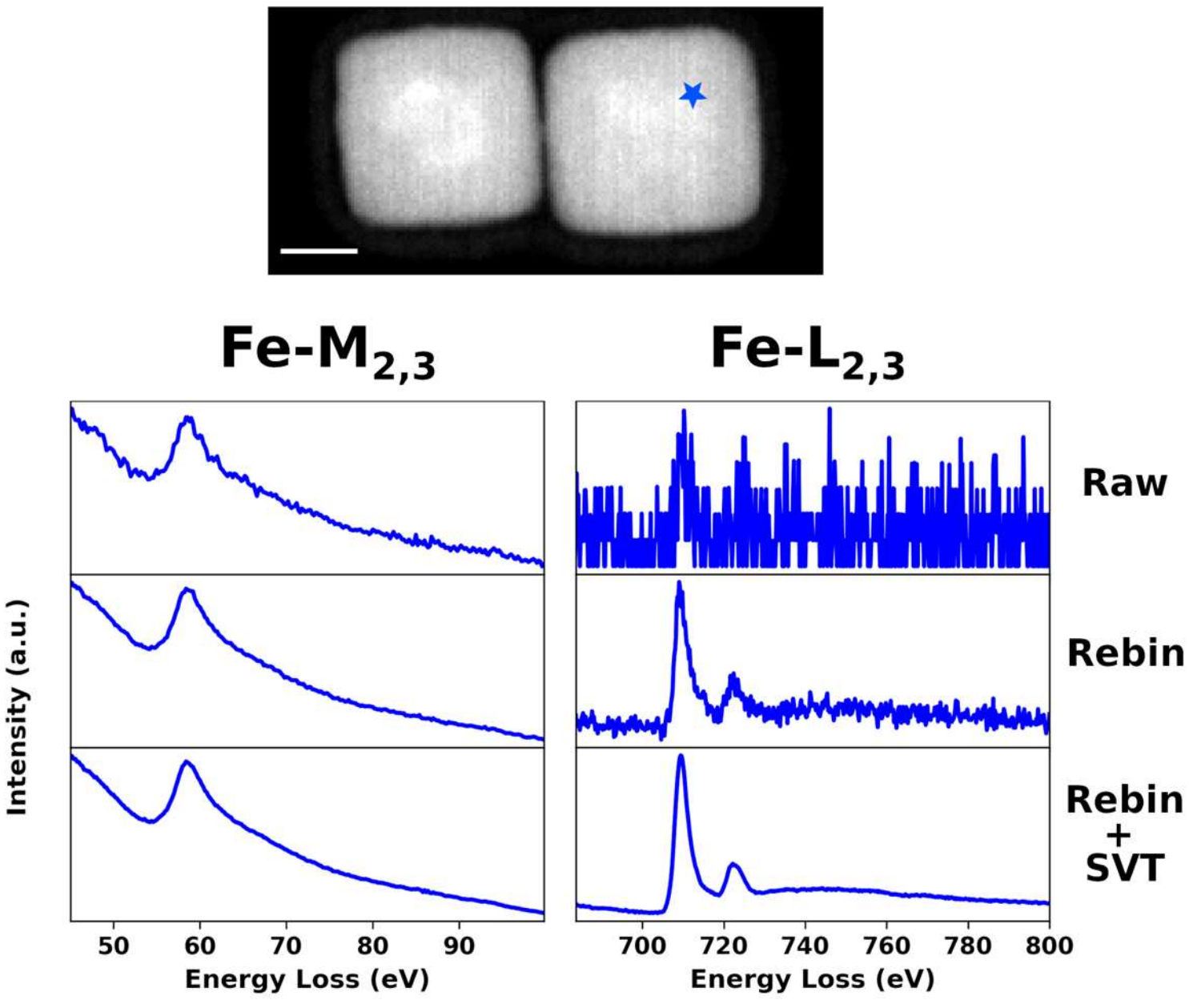

*Figure 1: Top: HAADF-STEM micrograph showing two of the FeO/$Fe_3O_4$ NPs. The area where all spectra has been taken from is highlighted with a blue star. Bottom: EELS intensity of raw spectra (top), spectra in 8x8 rebinned SPIMS (center) and SVT-denoised spectra (bottom) for Fe-$M_{2,3}$ and Fe-$L_{2,3,}$ edges, at equivalent dwell times. Measurement point is highlighted with a star. Scalebar is 10nm.*

Once the Fe-$L_{2,3}$ fingerprints had been found, finding their Fe-$M_{2,3}$ counterparts consisted of solving a linear equation system that expresses the fact that, at each point of the dataset, EELS analysis with the $L_{2,3}$ and $M_{2,3}$ edges must be equal. Indeed, the background-subtracted EELS experimental spectra can be expressed as a linear combination of the two fingerprints of their corresponding edge ($M_{2,3}$ at lower energies and $L_{2,3}$ at higher energies). The detailed process of these calculations is explained in the Supplementary Information, and their outcome, both spectral and in terms of abundance mapping, can be seen in Figure 2.

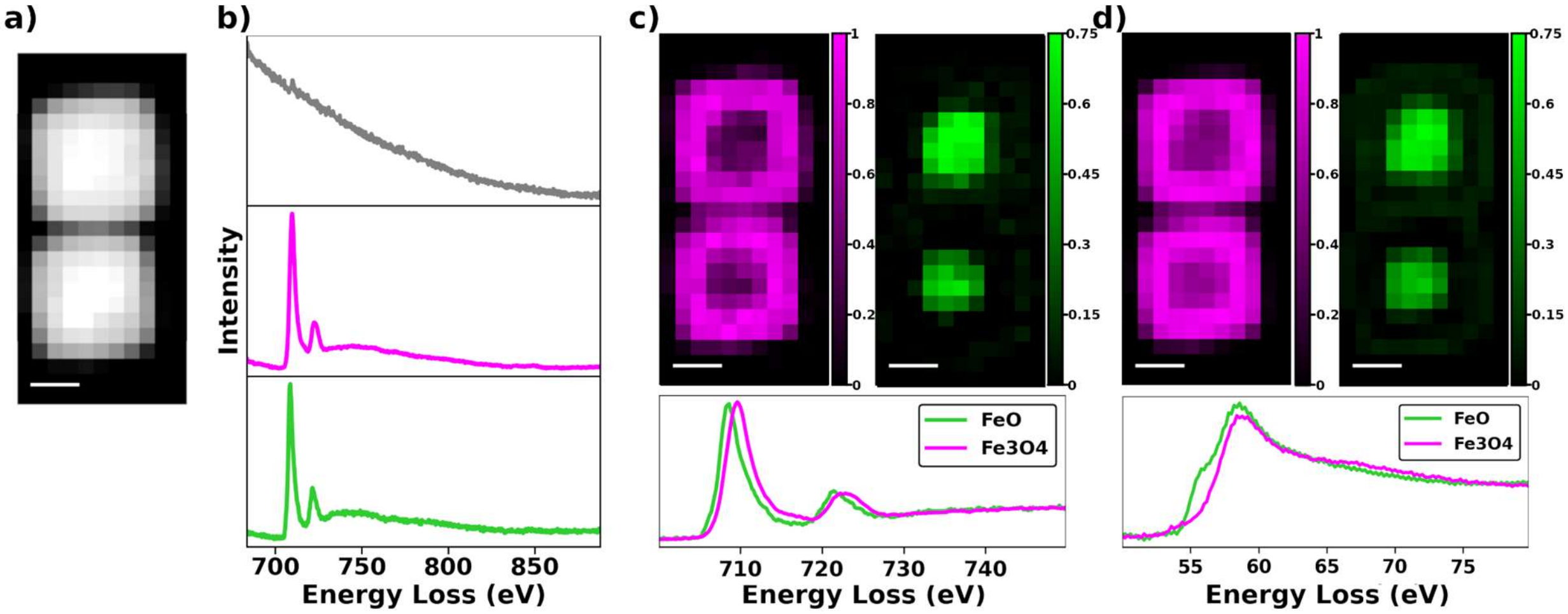

*Figure 2. a) HAADF intensity of the rebinned reference SPIMs. b) F-CL BSS components corresponding to $Fe_3O_4$ (magenta) and FeO (green). c) Spectra and maps of the Fe-$L_{2,3}$ references for $Fe_3O_4$ and FeO. Intensity is normalised by the maximum of the $Fe_3O_4$ map. d) Spectral comparison and mapping of the Fe-$M_{2,3}$ references for FeO and $Fe_3O_4$. Intensity is normalised by the maximum of the $Fe_3O_4$ map. All scalebars are 10nm.*

Once the Fe-$M_{2,3}$ spectral references for both FeO and $Fe_3O_4$ had been found, we can fit them to any LL dataset using a model consisting of the two fingerprints, a power law background and a convolution with the LL signal up to 40 eV to account for multiple-scattering (Manoubi et al., 1990; Verbeeck & Van Aert, 2004), from where we can directly obtain maps of the different iron species using only Soft EELS. We apply this method to our M-LL dataset, after performing a masked SVT denoising routine like the one mentioned before to stabilize the fit of the background on the narrow energy window (52:60 eV) in which we fit the model. As for the T-LL measurements, another 4D datacube (angle, x, y, ELoss) was created by combining the single frame acquisitions for all angles, and the same denoising and modelling routine was performed for the whole 4D datacube. The tilt-maps for FeO and $Fe_3O_4$ were aligned using the center of mass (CoM).

### Increase in dose efficiency

For the same dataset that has yielded the Fe-$M_{2,3}$ references in Figure 2, we have performed an estimation of the SNR and therefore of the dose efficiency of using Soft EELS for this kind of measurements.

The noise of our datasets follows a Poisson distribution, hence, its standard deviation is given by $\sqrt{C}$, with $C$ being the total amount of counts on a specific energy region. Therefore, the SNR of a specific energy window can be calculated as $SNR = \frac{S}{N} = \frac{C_S}{\sqrt{C}}$, with $C_s$ being the total counts of the background-extracted signal in the aforementioned energy region.

After applying this method to the previous dataset, we obtain that $\frac{SNR_M}{SNR_L} = 6.84$, that is, we get a 684% increase in the SNR when using the Fe-$M_{2,3}$ edges. In terms of the dose, if we were to multiply the L edge dose by a factor x, the SNR would be multiplied by a factor of $\sqrt{x}$, since $SNR_x = \frac{x \cdot C_s}{\sqrt{x \cdot C}} = \sqrt{x} \cdot \frac{C_s}{\sqrt{C}} = \sqrt{x} \cdot SNR_0$. We can then find the dose factor by which we would have to multiply the Fe-$L_{2,3}$ edge measurements to find the same SNR as the Fe-$M_{2,3}$ edge measurements:

$$\frac{SNR_M}{SNR_{L_x}} = 1 \Rightarrow \frac{SNR_M}{\sqrt{x} \cdot SNR_L} = 1 \Rightarrow x = \left(\frac{SNR_M}{SNR_L}\right)^2 = 46.72$$

In other words, we would need 46.72 times more dose to get the same SNR from the Fe-$L_{2,3}$ edges.

## DIPm-TV reconstruction method

In electron tomography, the simultaneous iterative reconstruction Technique (SIRT) is one of the most widely used algorithms for reconstructing 3D volumes from projection data by solving the following linear system:

$$y = Px + n \quad (1)$$

where $y$ denotes the acquired projections, $x$ denotes the unknown 3D object to reconstruct, $P$ is the projection operator (discretized 3D Radon transform), and $n$ accounts for measurement noise. The SIRT reconstruction method minimizes a least-squares ($l2$-norm) data-fidelity term that enforces consistency between the measured projections and the forward projections of the intermediate reconstructions:

$$\hat{x} = argmin_x \frac{1}{2}\left||Px - y|\right|_2^2 \quad (2)$$

When applied to limited-angle or sparse-view acquisitions, SIRT produces highly degraded reconstructions because the underlying inverse problem is ill-posed. In this context, a common alternative is compressed sensing (CS)-based approaches, which incorporate a regularization term promoting the sparsity of the object in a chosen transformation domain $L$ (i.e., gradient, wavelets, etc.) (Jacob et al., 2021; Leary et al., 2013; Saghi et al., 2016). This leads to the following regularized minimization problem:

$$\hat{x} = argmin_x \left\{\frac{1}{2}\left||Px - y|\right|_2^2 + \lambda \cdot \left||Lx|\right|_1\right\} \quad (3)$$

where the first term corresponds to the data-fidelity term, the second term corresponds to the regularization term, and $\lambda$ is a parameter that controls the trade-off between them.

In the present work, we target piecewise-constant reconstructions, which are well promoted by total variation (TV) regularization, defined as the $l1$-norm of the gradient: $TV(x) = \left||\nabla x|\right|_1$(Jacob et al., 2021). The resulting CS-TV reconstruction problem becomes:

$$\hat{x} = argmin_x \left\{\frac{1}{2}\left||Px - y|\right|_2^2 + \lambda \cdot \left||\nabla x|\right|_1\right\} \quad (4)$$

While CS-TV outperforms SIRT for moderately limited angle or sparse-view acquisitions, it exhibits limitations under extreme acquisition conditions, such as those employed in the present work.

In this work, we solve the data fidelity term in (4) by using the deep image prior (DIP) framework (Baguer et al., 2020; Brosset et al., 2026; Ulyanov et al., 2020) where the unknown 3D volume is parameterized by a convolutional neural network (CNN), $F_\theta\ (z)$, fed with a fixed random input $z$:

$$x = F_\theta(z)$$

Under this parametrization, the reconstruction is obtained by optimizing the network parameters $\theta$:

$$\hat{\theta} = argmin_\theta \left\{\frac{1}{2}\left||PF_\theta(z) - y|\right|_2^2 + \lambda \cdot \left||\nabla F_\theta(z)|\right|_1\right\} (5)$$

and the reconstructed object is given by $\hat{x} = F_{\hat{\theta}}(z)$. We name the approach DIP-TV.

In our implementation, $F_\theta$ is a 3D U-Net architecture, which, as an implicit prior, promotes local correlations and sharper, more coherent borders in the reconstructed volume, while TV regularization further enforces a piecewise-constant behaviour.

Moreover, we extend this DIP-TV approach in two directions. First, we replace the l2-norm in the data fidelity term with an l1-norm, chosen for its improved reconstruction quality on noisy data, as reported in (Barutcu et al., 2021). Second, we introduce a multi-channel formulation (DIPm-TV), in which the tilt series corresponding to different signals (in the present work, the oxidation state maps) are reconstructed jointly. This joint reconstruction leverages shared spatial information across channels, promoting inter-channel consistency and improving the reconstruction of interfaces. In particular, channels with lower SNR benefit from the structural support provided by higher-SNR channels, reducing channel-specific artifacts while preserving co-localized features. Importantly, this formulation also implicitly encourages a degree of inter-channel complementarity, effectively promoting solutions in which the reconstructed signals exhibit partially anti-correlated spatial distributions.

Let $y^c$ and $x^c$ denote the projections and volume of channels $c \in \{1, \cdots, N\}$, where $N$ is the total number of channels. We use a single 3D U-Net that outputs the parameters corresponding to all channels simultaneously, $F_\theta(z) = [F_\theta^1(z), \cdots, F_\theta^N(z)]$, and solve the following optimization problem:

$$\hat{\theta} = argmin_\theta \sum_{c-1}^{N} \quad \left\{\left||PF_\theta^c(z) - y^c|\right|_1 \ + \lambda \cdot \left||\nabla F_\theta^c(z)|\right|_1\right\} (6)$$

Note that for $N = 1$, (6) reduces to the standard single-channel DIP-TV formulation (Baguer et al., 2020; Ulyanov et al., 2020). After optimization, the reconstructed multi-channel volume is expressed as:

$$\hat{x} = [\hat{x}^1, \cdots, \hat{x}^N] = \left[F_{\hat{\theta}}^1(z), \cdots, F_{\hat{\theta}}^N(z)\right]. \quad (7)$$

SIRT reconstructions are computed using the ASTRA Toolbox (Van Aarle et al., 2015, 2016), CS-TV reconstructions are performed with pysap-etomo (Farrens et al., 2020), the forward projector is implemented using Tomosipo (Hendriksen et al., 2021), and optimization is performed in PyTorch.

For DIPm-TV, the learning rate, the noise regularization strengths, the TV weight ($\lambda_{TV}$), and the architecture of the 3D U-Net were selected from previous works (Baguer et al., 2020; Barutcu et al., 2021; Ulyanov et al., 2020). We note that no universal criteria are available to date for determining the optimal hyperparameters in an automated way (Benfenati et al., 2025; Wang et al., 2021). The number of training iterations is selected based on the loss function and visual assessment of intermediate reconstructions to avoid overfitting-induced artifacts through early stopping.

### DIPm-TV applied to simulated data

In this work, EELS tomography was performed using sparse angular sampling, with a tilt increment of 17.5° over a wide angular range from -70° to +70°, corresponding to a missing-wedge of 40°. To quantitatively assess the performance of DIPm-TV under these acquisition conditions, we constructed a synthetic phantom replicating the core-shell morphology of the nanocube (Figure 3a). This controlled setup allows us to isolate the effects of the acquisition geometry and noise, avoiding additional uncertainties inherent to experimental data. To this end, we generated a simulated tilt-series dataset of the core-shell structure using the same acquisition geometry as in the experiment: 9 projections were generated from -70° to +70°, with a 17.5° tilt increment. Poissonian noise was then added to emulate various low-dose conditions.

Figure 3a compares SIRT, CS-TV and DIPm-TV reconstructions of the simulated core-shell phantom at an SNR of 12.7 dB. The SIRT reconstruction exhibits artifacts arising from the large tilt increment, the missing wedge (displayed vertically), and the low-SNR conditions. As a consequence, the reconstruction appears noisy and distorted, and the thin outer shell cannot be recovered. CS-TV significantly improves the reconstruction quality by reducing noise and mitigating sparse-view and missing-wedge artifacts for the core and shell. Nevertheless, it still fails to recover the thin outer shell. This limitation arises from the use of TV as a unique sparsifying transform, which promotes piecewise-constant structures throughout the reconstruction. While this assumption is suitable for the core and shell regions, it is not well adapted to the curve-like features associated with the thin outer shell. Therefore, for structures containing multiple feature types, a single sparsifying transform cannot efficiently represent all morphological components simultaneously, showing an inherent limitation of CS-based approaches. In contrast, DIPm-TV reconstructs the core-shell morphology with high fidelity, reliably resolving the thin outer shell even along the missing-wedge direction. This demonstrates its ability to preserve fine structural details under highly sparse-view and low-dose conditions.

In this context, Figure 3b shows the horizontal intensity profile of the core signal, comparing CS-TV and DIPm-TV against the reference. This profile highlights the capability of DIPm-TV to closely match the reference, accurately recovering the outer shell thickness, whereas CS-TV fails to resolve it and produces a nearly uniform signal across the region. Importantly, denormalized DIPm-TV profiles preserve near-absolute signal levels, indicating that the method remains suitable for quantitative analysis beyond qualitative morphological recovery. A more extensive evaluation across different noise levels is provided in the Supplementary Information.

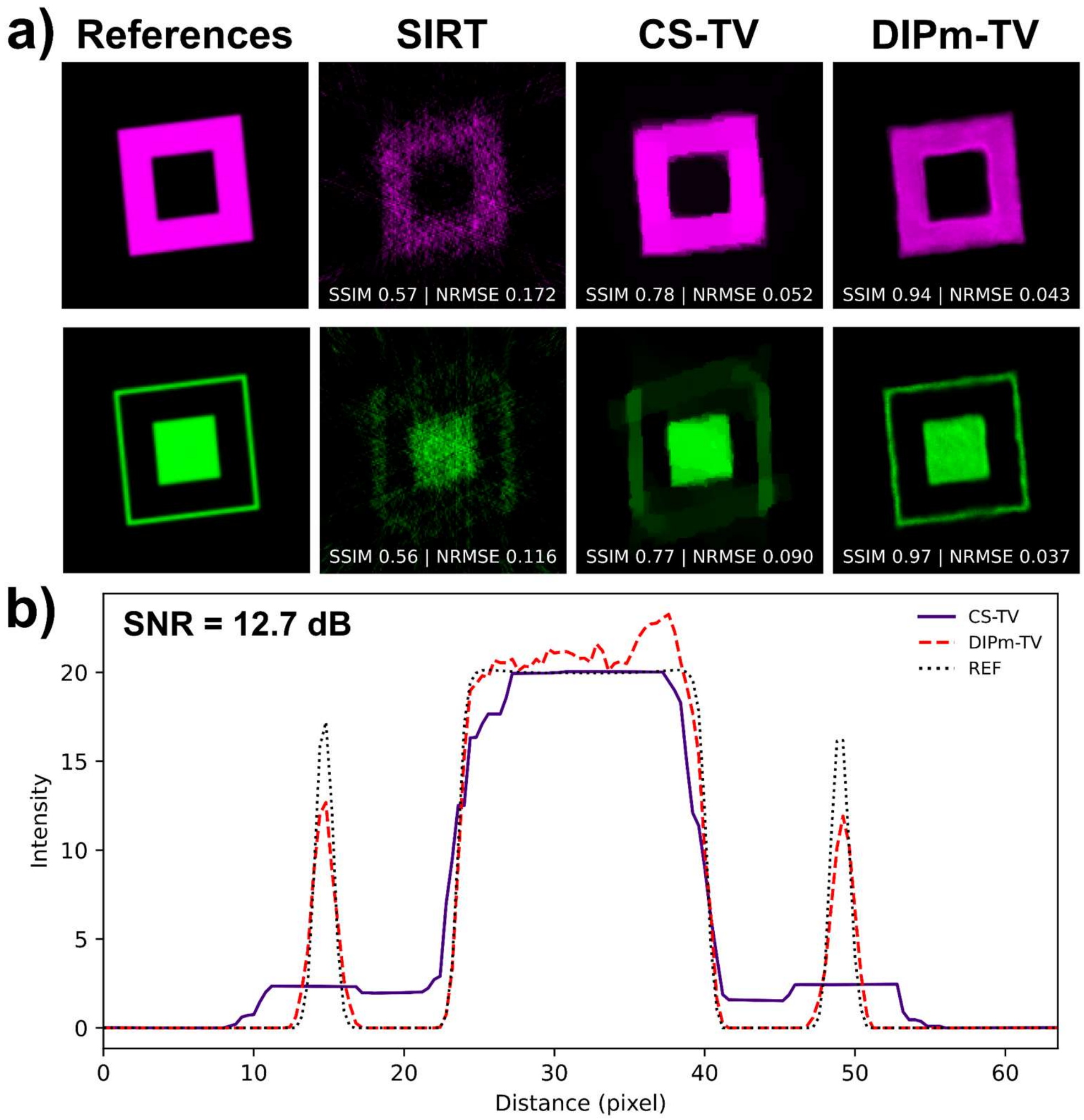

*Figure 3.* Reconstruction methods applied to the simulated core-shell phantom at an SNR of 12.7 dB, under the same angular conditions as the EELS tomography experiment [-70°:17.5°:+70°]. (a) From left to right, central cross-section views of the core-shell phantom reconstructed using: reference (SIRT reconstruction from [-90°:1°:+90°]), SIRT, CS-TV and DIPm-TV. Each view also includes quantitative volumetric metrics (SSIM and NRMSE). (b) 1D horizontal line profiles from the central slice of the core channel, comparing CS-TV and DIPm-TV. The reference (REF) is shown as a black dotted line.

# Results and discussion

## *2D oxidation state mapping*

Figure 4 displays the abundance maps for FeO and $Fe_3O_4$ derived from Fe-$M_{2,3}$ ELNES analysis, showing features that are just a few pixels wide (pixel size is 0.4 nm).

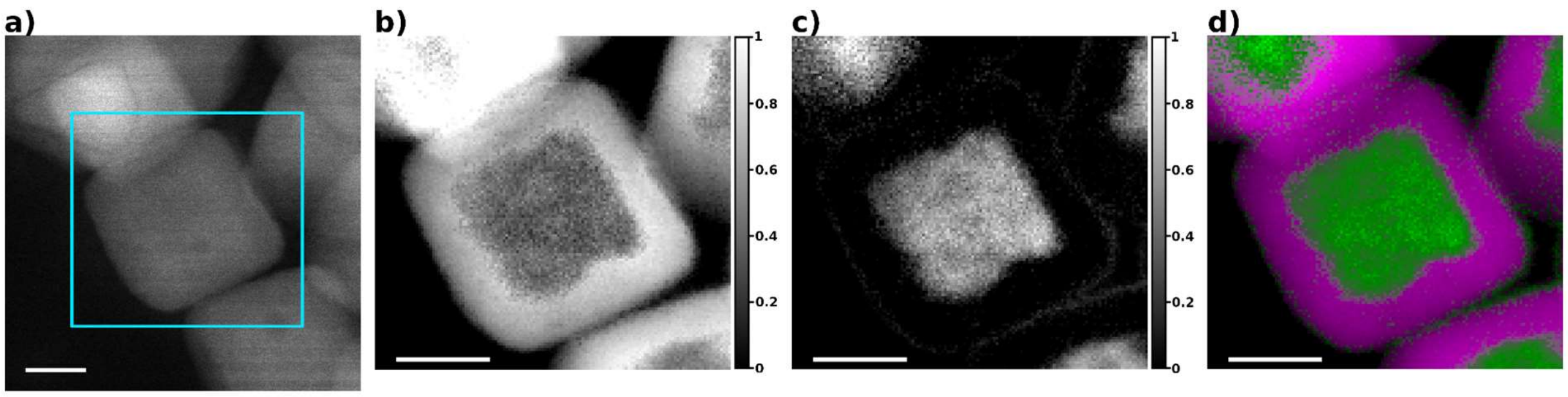

*Figure 4. Fe-$M_{2,3}$ mapping of FeO and $Fe_3O_4$. a) HAADF micrograph of a small nanocube cluster. The highlighted area shows the region in which the SPIM was taken. b) Abundance maps of $Fe_3O_4$ and c) FeO from the analysis of the Fe-$M_{2,3}$ ELNES features. Intensity has been normalised with respect to the maximum value of the $Fe_3O_4$ intensity in the central NP. d) Overlay image of the $Fe_3O_4$ and FeO ELNES maps showing a core-shell-shell spatial distribution. Scale bars are all 20 nm.*

The HAADF image (Figure 4a) shows a c.a. 50 nm nanocube surrounded by other partially overlapping particles. The core seems to be slightly brighter than the shell, suggesting that it is made of FeO, whose HAADF intensity is slightly higher than $Fe_3O_4$. The nanocube is surrounded by a 2-3nm layer that we attribute to a byproduct of the organic ligands in the nanoparticle synthesis and dispersion (C. Li et al., 2021). Even though the sample (but not the particle) is the same as in (Torruella et al., 2016), it has undergone apparent changes in the 10-years gap between the experiments. Specifically, the small rounded dark features in the HAADF image suggest the presence of small spherical voids in the sample. This has been previously reported in similar materials, and attributed to the Kirkendall effect, due to the difference in oxygen and iron diffusion between FeO and $Fe_3O_4$ (Cabot et al., 2007; Sun et al., 2017). These voids were not reported in (Torruella et al., 2016), however we see them from the very beginning of these experiments. We conclude that these can be attributed to the aging of the sample during the years.

Once the ELNES maps have been obtained, and the core-shell nature has been properly shown, it seems the $Fe_3O_4$ shell of the nanoparticle (Figure 4b) appears to be thicker than in (Estrader et al., 2015; Torruella et al., 2016). This implies that the surface of FeO cores has been slowly oxidizing onto $Fe_3O_4$, going from a shell measuring 8-9 nm in previous studies to 9-12 nm in this one. This is coherent with the finding of Kirkendall-caused voids, and they both prove the migration of iron and oxygen between the core and the shell of the particle. However, much like in the bibliography, these maps have not found any degree of alloying between the FeO core and the $Fe_3O_4$ shell. From an analytical point of view, the diminishing of the core further complicates the ELNES analysis of the nanoparticles, decreasing the SNR in the core both due to its lower thickness and to the higher thickness of its surrounding shell.

The last feature of these NP found on the ELNES maps, and one of special interest for not being reported before in the literature, is a very thin (approximately 1 nm thick) layer of FeO on the outermost part of the nanoparticle (Figure 4c). It is not clear yet how this layer has come to be. Several hypotheses for the formation of this outer layer include the ultra-high vacuum inside the electron microscope, electron irradiation, the thermal treatment the sample was subjected to before entering the microscope, reduction coming from the surrounding carbon layer; as well as a combination of these four possible factors. The first two hypotheses are of special interest because they would have been present in previous studies on these samples, but they would have only been visible now with the heightened spatial resolution and SNR provided by Soft EELS. This layer is present in all the NPs we have mapped and analyzed. This means that, if irradiation were to play a role in its formation, it would only need dose lower that the one needed for a single map ($\sim 2.3 \cdot 10^4$ $e^- \cdot Å^{-2}$). In any case, it is clear that the aforementioned pollution layer plays a crucial role in preserving this FeO thin layer. When exposed to air, the surface of FeO rapidly transforms into $Fe_3O_4$, which would make this conformation unstable unless it was somehow shielded from air (Sharma et al., 2011).

This thin layer had not been reported before in the literature to the best of our knowledge, either because it did not exist at the time of the previous experiments, or because the techniques at the time did not have a sufficient SNR to be able to discern it. Its conformation is also a mystery, not knowing whether it is a true thin film or just a series of islands on the NP surface. The very thin nature of this layer, along with the presence of the original core and shell features of the nanoparticle, makes this an impossible feat for 2D mapping. Similarly, we can know of the presence of the Kirkendall-caused voids, but it is impossible to locate them within the NP. For these reasons, the only possible next step in order to solve these questions is to extend this analysis to 3D.

### 3D oxidation state mapping

An isolated nanoparticle was selected to perform the tilt-series acquisition shown in the T-LL dataset. In Figure 5, we show the HAADF-STEM micrographs, as well as the centered ELNES maps corresponding FeO (green) and $Fe_3O_4$ (magenta) signals for each acquisition angle are displayed in Figure 5.

In order to perform the 3D reconstructions to the full extent of their potential, a particularly challenging NP has been chosen as an example..Its small size, combined with the more severe oxidation the sample has undertaken over the years, has resulted in a particularly small core. A first approximation from the ELNES maps at 0º estimates the core of this specific NP only represents 10% of the total volume, making the ELNES analyses, as well as the reconstruction, even more difficult.

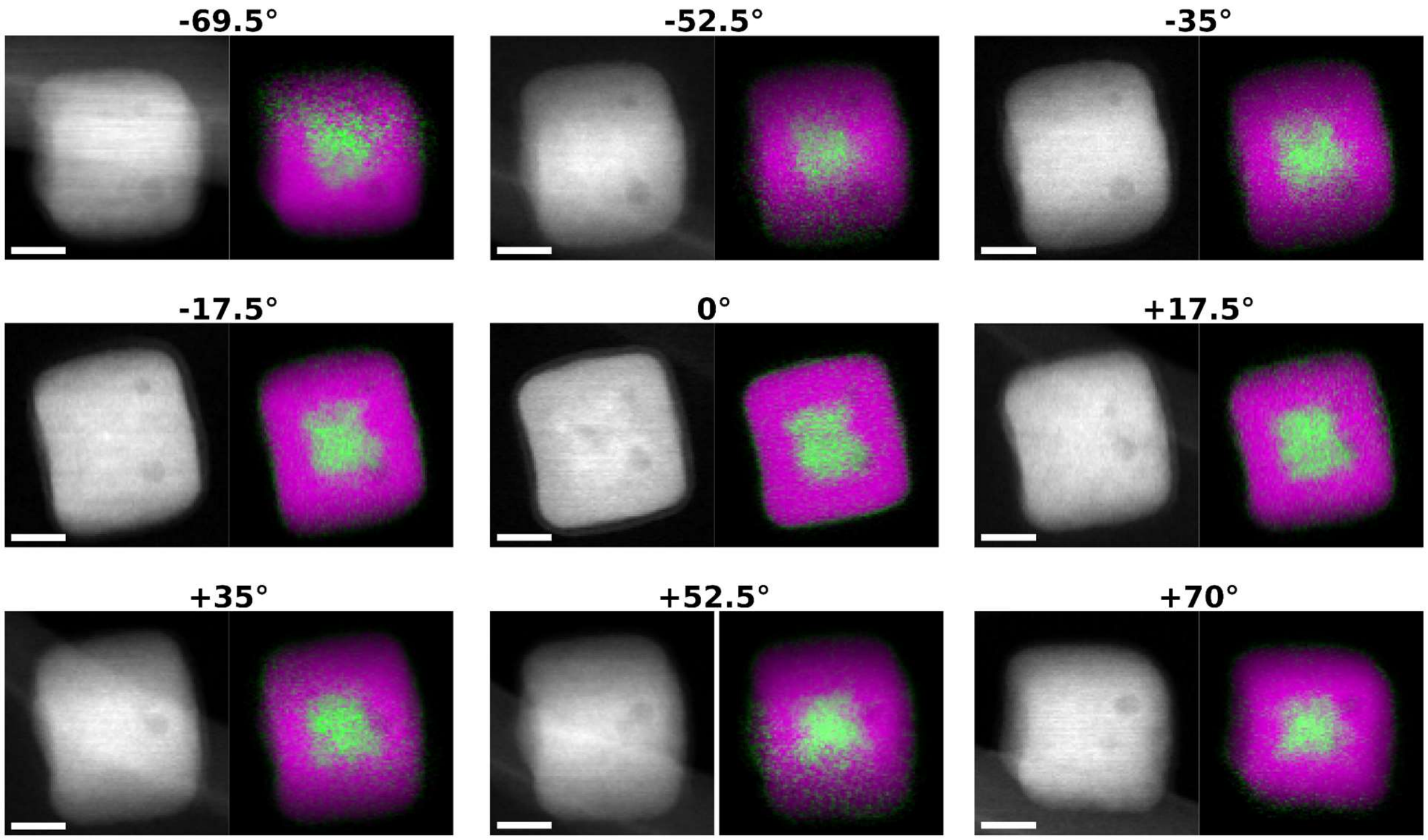

*Figure 5: HAADF-STEM (left) and ELNES maps taken from the Fe-$M_{2,3}$ edges (right, FeO in green and $Fe_3O_4$ in magenta) for each of the 9 tilting angles in the tilt-series. Scale bar is 10 nm.*

Furthermore, the location of the nanoparticle within the carbon membrane on a very thin strip further complicates this study. In a homogeneous grid, the thickness of carbon membrane passed through by the electron beam at a tilt angle alpha ($t_\alpha$) varies with the tilt angle as $t_\alpha = \frac{t_0}{cos(\alpha)}$, where $t_0$ is the thickness of the membrane. This increase in thickness (which is almost tripled at the highest values of $\alpha$ in these experiments) homogeneously lowers the SNR of the ELNES maps at these angles due to multiple scattering.

However, the position of the NP within the carbon membrane results in its partial projection over two different holes in the membrane for several tilt angles, providing different levels of SNR within the same map.

Even in these extremely difficult conditions, the ELNES analyses have been able to retrieve features related to the Kirkendall voids in this particular NP, as well as the reduced FeO layer at the very edge of the particle, which can be discerned at angles from -35º to +17.5º. This shows the potential of Soft EELS for the identification of oxide phases, even at particularly challenging conditions.

Regarding the tomographic reconstructions of these ELNES maps, the renderings from the CS-TV and DIPm-TV reconstructions can be seen in Figure 6. The CS-TV volume exhibits noise and speckle-like artefacts throughout the reconstructed nanocube, whereas DIPm-TV suppresses most of these artefacts, resulting in a cleaner reconstruction. Moreover, the outer shell appears more continuous in DIPm-TV reconstructions than in CS-TV reconstructions. This visual improvement is further supported by orthogonal slices through the reconstructions.

In the XY cross-section, the only one with no missing wedge in Figure 6c-f, CS-TV shows a less reliable reconstruction of the external FeO shell, which appears much thicker and almost completely disappears in the top part of figure 6c. The external shell is not only fully retrieved using DIPm-TV, but furthermore its thickness is much more coherent with what is seen in the ELNES projections. CS-TV also yields noisier cross-sections, with higher small-scale intensity fluctuations across both the core and shell maps.

The advantages of DIPm-TV as a reconstruction strategy are even more evident in the YZ views (Figure 6i–j) and XZ views (Figure 6m–n) which include the missing wedge direction (displayed vertically). In these cross-sections, CS-TV reconstruction (Figure 6g-h) presents star- and streaking-like artefacts due to the sparse-view sampling (Jiang et al., 2018), visible as diagonal and linear low-intensity lines in Figure 6h. In addition, a shadowing contribution from the carbon support grid appears as a localized distortion of the core signal (top region of Figure 6g and 6k). Also, even though the missing wedge is mild, its effect is stronger for curve-like structures such as the thin outer shell. As a result, CS-TV does not recover the horizontal structure of the thin outer shell, as expected from the simulations. In contrast, DIPm-TV yields more isotropic and spatially coherent reconstructions, effectively recovering fine structural details in all 3 directions, as shown in Figures 6e, 6i and 6m. We note that the DIPm-TV reconstruction of the thin outer shell is not uniform, which can be due to heterogeneity of this FeO shell. In addition, we should notice that this thin outer shell is not visible in all the FeO maps, due to the low SNR.

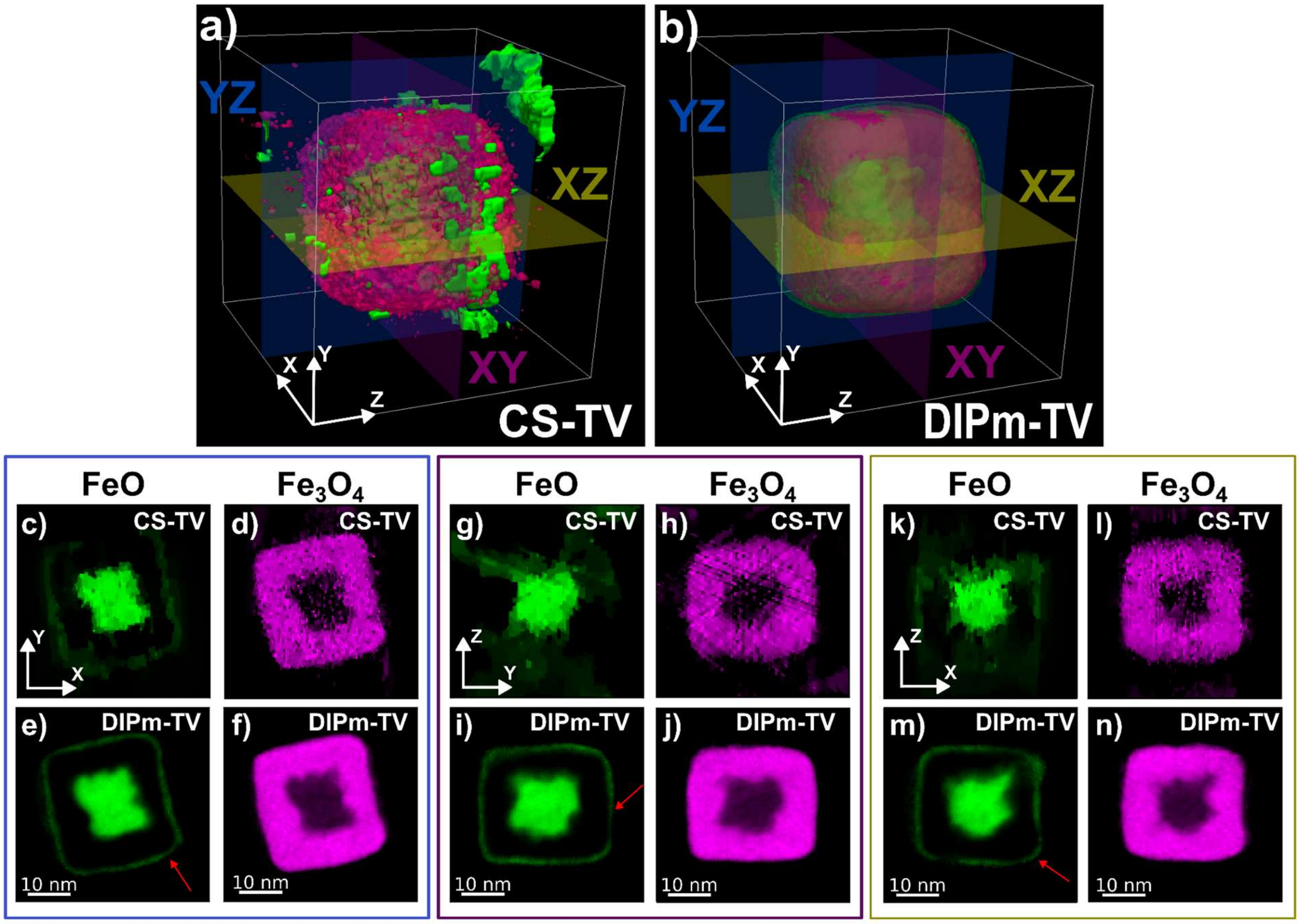

*Figure 6. Central orthogonal slices and 3D volume renderings of the reconstructed EELS oxidation-state maps obtained by STEM-EELS tomography using CS-TV and DIPm-TV. For each reconstruction method, central cross-sections are shown in the XY, YZ, and XZ planes for the core (green) and shell (magenta) oxidation-state maps. (a-b) display 3D volume renderings for a representative conventional reconstruction (CS-TV) and the DIPm-TV. (c–f) presents the XY cross sections view, (g-j) displays the YZ cross sections view and (k-n) shows the XZ cross sections view. In this reconstruction, the missing-wedge direction is aligned with the vertical axis, leading to anisotropic resolution primarily along Z.*

In Table 1, the spatial resolution achieved for each reconstruction algorithm is estimated by the power spectral density (PSD)-derived cutoffs. SIRT exhibits the strongest degradation along the missing-wedge direction, with the largest XY-to-Z anisotropy ratio (1.38). CS-TV partially mitigates this effect, reducing the Z anisotropy (1.26) and indicating an improved recovery of high spatial frequencies along Z compared to SIRT. Finally, DIPm-TV yields the most isotropic behavior, with near-unity ratios (0.97), consistent with a strong suppression of missing-wedge-induced anisotropy and a more uniform frequency support across all 3 directions.

Beyond anisotropy, Table 1 also indicates the effects of SNR imbalance between signals: for both SIRT and CS-TV, the $Fe_3O_4$ shell achieves better cutoff resolutions than the FeO core, consistent with the lower effective SNR in the core due to multiple scattering. Notably, DIPm-TV compensates for this SNR imbalance thanks to the multichannel approach, delivering comparable resolutions for both core and shell.

| Method/Signal | Resolution XY (nm) | Resolution Z (nm) | Difference (nm) | Ratio |
|---|---|---|---|---|
| **SIRT FeO** | 1.36 | 1.88 | 0.51 | 1.38 |
| **SIRT $Fe_3O_4$** | 1.24 | 1.75 | 0.51 | 1.38 |
| **CS-TV FeO** | 1.32 | 1.66 | 0.34 | 1.25 |
| **CS-TV $Fe_3O_4$** | 1.23 | 1.60 | 0.37 | 1.28 |
| **DIPm-TV FeO** | **1.04** | **0.99** | **0.05** | **0.95** |
| **DIPm-TV $Fe_3O_4$** | **0.91** | **0.91** | **0.00** | **1.00** |

*Table 1. Power spectral density-based spatial resolution for the Fe oxidation-state volumes. The lateral (XY) resolution is reported as the mean of the cutoff resolutions obtained along $k_X$ and $k_Y$, while the axial (Z) resolution corresponds to the cutoff along $k_Z$ (missing-wedge direction). The difference denotes $\Delta = Res_Z - Res_{XY}$ and the ratio denotes the $Res_Z/Res_{XY}$.*

The recovery of the thin shell can be better appreciated in Figure 7, where vertical intensity profiles of the central YZ cross-sections are shown. In Figure 7a, the CS-TV profile exhibits clear staircasing artefacts in the FeO core and $Fe_3O_4$ shell profiles, which is typical of TV regularized reconstructions (Jiang et al., 2018). In addition, there is no clear trace of the thin outer shell. In contrast, the DIPm-TV profile is smoother and the thin outer shell is recovered (red arrows in Figure 7).

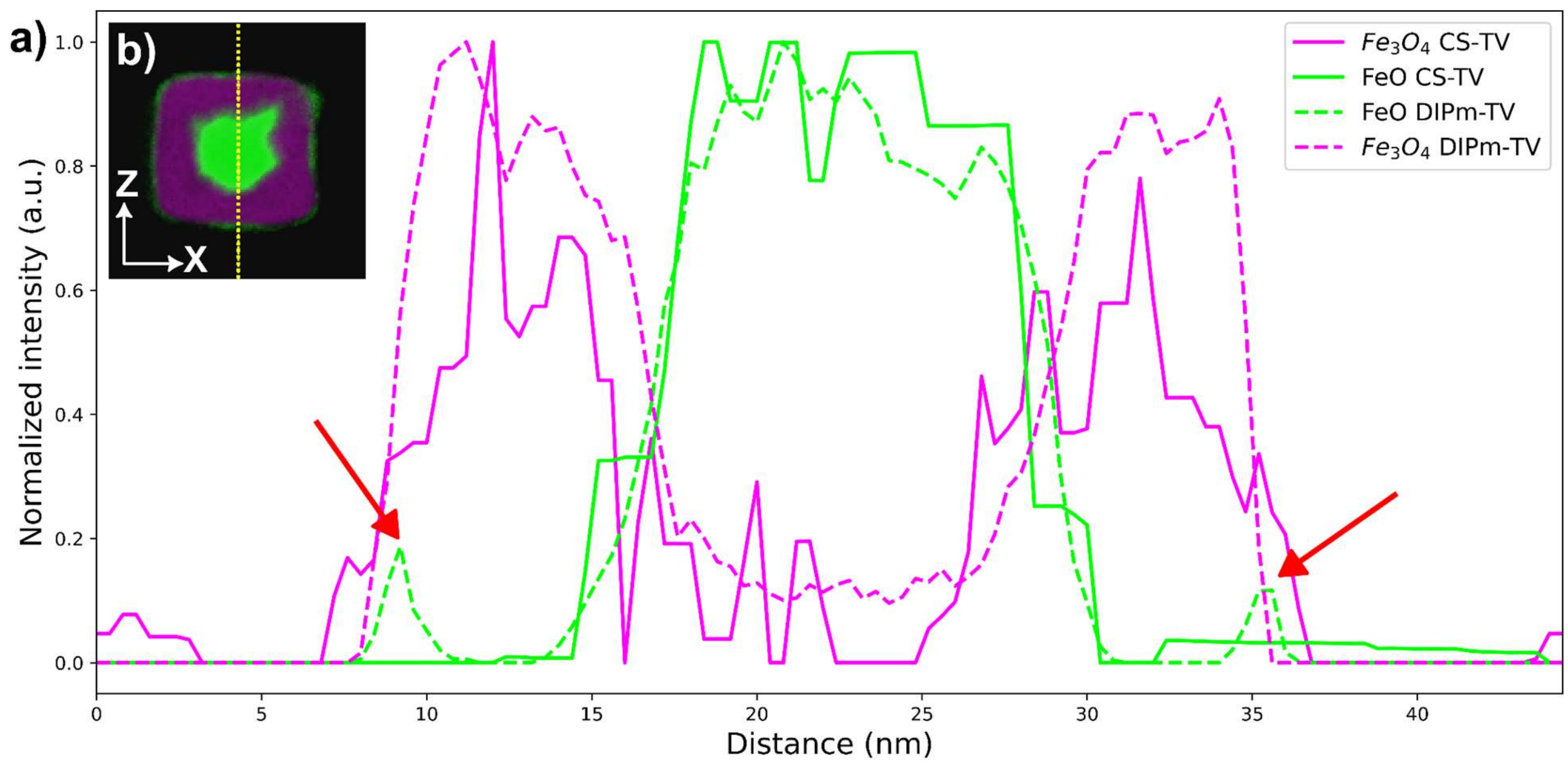

*Figure 7. (a) Vertical intensity profiles from CS-TV and DIPm-TV reconstructions. (b) XZ cross-section view from DIPm-TV with the corresponding profile (yellow dashed line). The $Fe_3O_4$ signal is shown in magenta and the FeO signal in green.*

In agreement to what has been seen in the 2D analyses, the 3D studies have also found voids within the core-shell nanoparticles, as shown in Figure 8. The void is clearly visible in shell reconstructions obtained with SIRT, CS-TV and DIPm-TV. However, CS-TV and SIRT exhibit residual signals within the void region, which partially obscures and distorts its boundaries, whereas DIPm-TV suppresses this noise and recovers the void location more accurately. However, in the core reconstruction, a non-zero intensity persists inside the void across all reconstruction methods. This is due to a noise-driven artefact that remains difficult to fully eliminate under these low-dose and highly sparse-view conditions. In addition, the voids are located near the core–shell interface, consistent with their origin from the Kirkendall effect caused by diffusion between the two oxides.

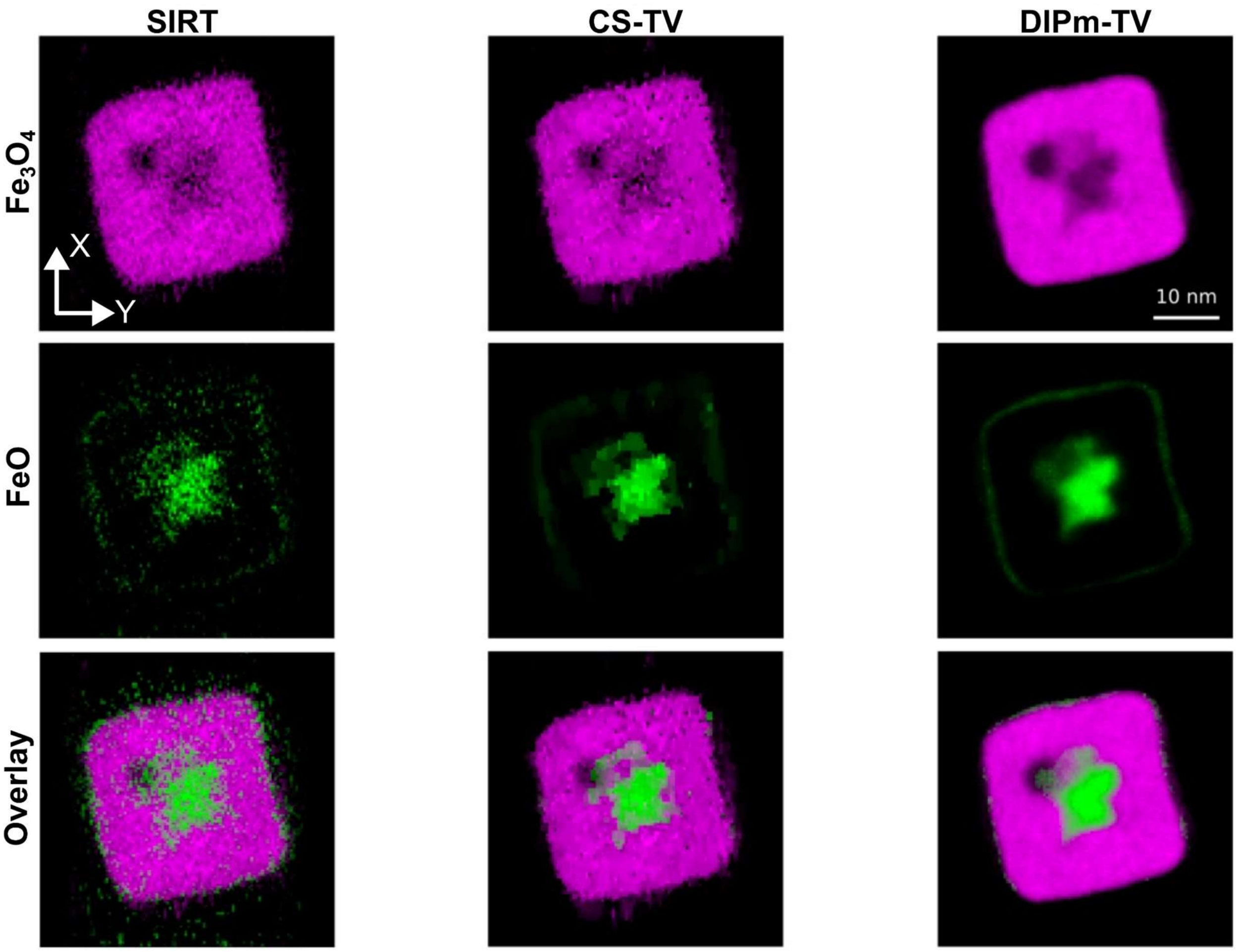

*Figure 8: Cross-section view visualizing the internal void within the shell and core reconstructions for CS-TV and DIPm-TV.*

## Conclusions

The two main developments combined in this study have excelled in reaching their objectives for the advanced characterization of Fe oxidation states. First, Soft EELS has provided a seven-fold leap in SNR by using the Fe-$M_{2,3}$ edges for bond mapping instead of the conventionally used Fe-$L_{2,3}$ edges. The complexity of the Fe-$M_{2,3}$ energy window in EELS has been overcome by previously finding reliable references, by correlating the intensity of the Fe-$M_{2,3}$ edge to that of the Fe-$L_{2,3}$ edges for a given calibration nanoparticle. The results have shown ELNES maps with a much higher resolution in FeO/$Fe_3O_4$ nanoparticles than anything reported in the literature, allowing the mapping of features with a comparable size to that of the electron probe. This bond mapping has been proven to be effective even in tomographic measurements at challenging conditions, due to the better SNR offered by soft EELS but also the nature of hybrid pixel detectors, which have allowed for the analysis of the maps taken at every angle at once, as a 4D dataset.

Second, we demonstrate the potential of DIPm-TV for 3D oxidation-state mapping in STEM-EELS tomography, under highly sparse-view acquisition (9 projections) and low-dose conditions. Unlike state-of-art strategies such as CS or MM frameworks, our approach does not impose symmetry constraints and does not rely on external priors such as model-based approaches or auxiliary HAADF signals. Instead, DIPm-TV provides an unsupervised framework that leverages only the acquired EELS projections to recover 3D oxidation-state volumes with nearly isotropic resolution, while achieving strong noise suppression thanks to the multichannel formulation. PSD-based directional analysis further shows that DIPm-TV outperforms conventional methodologies in terms of effective resolution, reaching approximately 1 nm resolution using only 9 projections.

Overall, these results open a new set of possibilities for the characterization of Fe-containing dose-sensitive materials with astounding spatial resolution, a critical tool for future iron oxide nanomaterial research. They also serve as a proof-of-concept for Soft EELS as a technique, opening the door to the advanced characterization of materials containing other elements in these Soft EELS energy ranges. This characterization is not limited to the study of beam-sensitive materials, but rather its much higher SNR and therefore faster acquisition times makes it a perfect candidate for applications where speed is a key factor, such as *in-situ* experiments. These new possibilities exceed even the realm of EELS, with the multichannel strategy introduced in this paper being also applicable to STEM-EDX tomography for 3D elemental mapping. Beyond sparse-view tomography, the methodology is readily applicable to limited-angle acquisitions, opening new possibilities for fast in-situ 3D characterization using dedicated tomography holders. Nevertheless, further work will be needed to be able to create a robust methodology to make Soft EELS an analysis routine.

# Acknowledgements

The author M.P.F thanks the funding by the Spanish University Ministry and Next Generation EU programs through a Margarita Salas Fellowship. R.A and M.P.F. acknowledge funding from the Spanish MICIU (PID2023-151080NB-I00/AEI/10.13039/501100011033 and CEX2023-001286-S MICIU/AEI /10.13039/501100011033), as well as from the Government of Aragon (DGA) through the project E13 23R. This experimental work was carried out on the electron microscopy facility of the Advanced

Characterization Platform of the Chevreul Institute, University of Lille—CNRS. This study has been funded by ISITE ULNE and the “Métropole Européenne de Lille” through the “TEM-Aster project”.

Authors S.E and F.P. acknowledge the Spanish Project PID2022-138543NB-C21 funded by MCIU/AEI/10.13039/501.10001103. M. E acknowledges grants PID2024-161653NB-I00 and CNS2024-154589 funded by MICIU/AEI/10.13039/501100011033. The tomographic reconstruction work, carried out on the Platform for Nanocharacterisation (PFNC) in CEA-Grenoble, was supported by the “Recherche Technologique de Base” program of the French ministry of research. The authors also acknowledge the French National Research Agency through the PEPR DIADEM.

# Supporting Information: Low-Dose 3D Bonding Mapping Through "Soft" Core-Loss EELS Tomography and Unsupervised Deep Learning

Mario Pelaez-Fernandez[1,2*], Daniel del-Pozo-Bueno[3], Adrien Teurtrie[1,4], Serge Brosset[3], Maya Marinova[5], Phillipe Ciuciu[7,8], Marta Estrader[9,10], Germán Salazar-Alvarez[11], Francesca Peiró[10,12], Raul Arenal[2,3], Sonia Estradé[10,12], Zineb Saghi[4], Francisco De la Peña[1*]

[1] Unité Matériaux et Transformations (UMET UMR 8207), Université de Lille, Bâtiment C6, Cité Scientifique, Villeneuve d'Ascq, 59650, France

[2] Instituto de Nanociencia y Materiales de Aragon (INMA), CSIC-U. de Zaragoza, Calle Pedro Cerbuna 12, Zaragoza, 50009, Spain

[3] Univ. Grenoble Alpes, CEA, Leti, F-38000 Grenoble, France

[4] Centre d'Elaboration de Matériaux et d'Etudes Structurales (CEMES), University of Toulouse and CNRS, 31055 Toulouse, France

[5] Univ. Lille, FR 2638-IMEC-Institut Michel-Eugène Chevreul, F-59000 Lille, France

[6] CEA, Joliot, NeuroSpin, Gif-sur-Yvette Cedex 91191, France.

[7] Inria, MIND, Université Paris-Saclay, Palaiseau 91120, France.

[8] Department of Inorganic and Organic Chemistry, Inorganic Chemistry Section, University of Barcelona, Carrer de Martí i Franquès, 1-11, 08028 Barcelona, Spain

[9] Institute of Nanoscience and Nanotechnology (IN2UB), Universitat de Barcelona, 08028 Barcelona, Spain.

[10] Department of Materials Science and Engineering, Ångström Laboratory, Uppsala University, Uppsala 751 03, Sweden

[11] LENS-MIND, Departament d'Enginyeria Electrònica i Biomèdica, Universitat de Barcelona, 08028, Barcelona, Spain

[12] Laboratorio de Microscopias Avanzadas, Universidad de Zaragoza, Calle Mariano Esquillor, Zaragoza, 50018,

Spain
*emails: mariopf@unizar.es, francisco.de-la-pena-manchon@univ-lille.fr

## 1. Use of ICA in Fe-$M_{2,3}$ edges and SVD results

An initial strategy for the discerning of the FeO and $Fe_3O_4$ ELNES features in the Fe-$M_{2,3}$ edges was employing a similar analysis routine to the one shown in the main text for the Fe-$L_{2,3}$ edges, consisting of a denoising via SVD and the implementation of BSS using 3 components for the background and each Fe species. The factors resulting from this initial decomposition can be seen in Figure S1:

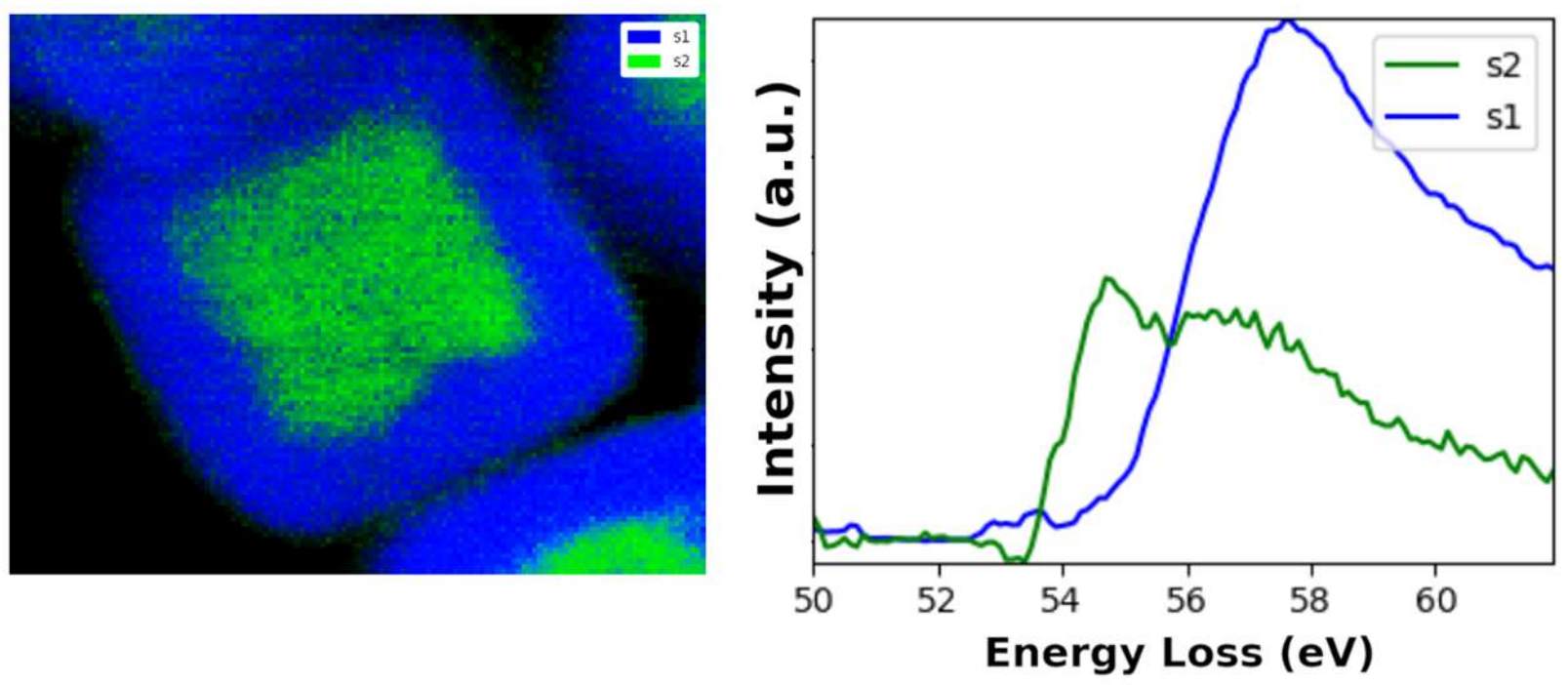

Figure S1: initial results of applying the BSS routine shown in the main text to the Soft-EELS signal. Results show that, while this preliminary analysis showed two distinct components and was able to separate the core and shell regions, these components did not match the Fe-$M_{2,3}$ fingerprints found in the study.

Even though the factors from this decomposition already hinted at the qualitative difference in the ELNES features, the spectral fingerprints shown in Figure S1 were not matching the Fe-$M_{2,3}$ spectra of similar materials in the literature (van Aken & Woodland, 1999). This was particularly visible in the case of the FeO in the core. Furthermore, analyses showed the ratio between the two components was not coherent with their Fe-$L_{2,3}$ counterpart.

It was thus concluded that the BSS method employed in the F-CL dataset could not be applied directly to the F-LL dataset. This is most probably due to the complexity of the signal in this energy range, which leads to a dataset that has a higher dimensionality that the number of chemical species, complicating BSS; as well as both components sharing features in the 56-60 eV energy, and the general non-linearity of this energy range.

As for the F-CL dataset, SVD results, illustrated through their corresponding Scree plot, which can be seen in figure S2, show three clear components.

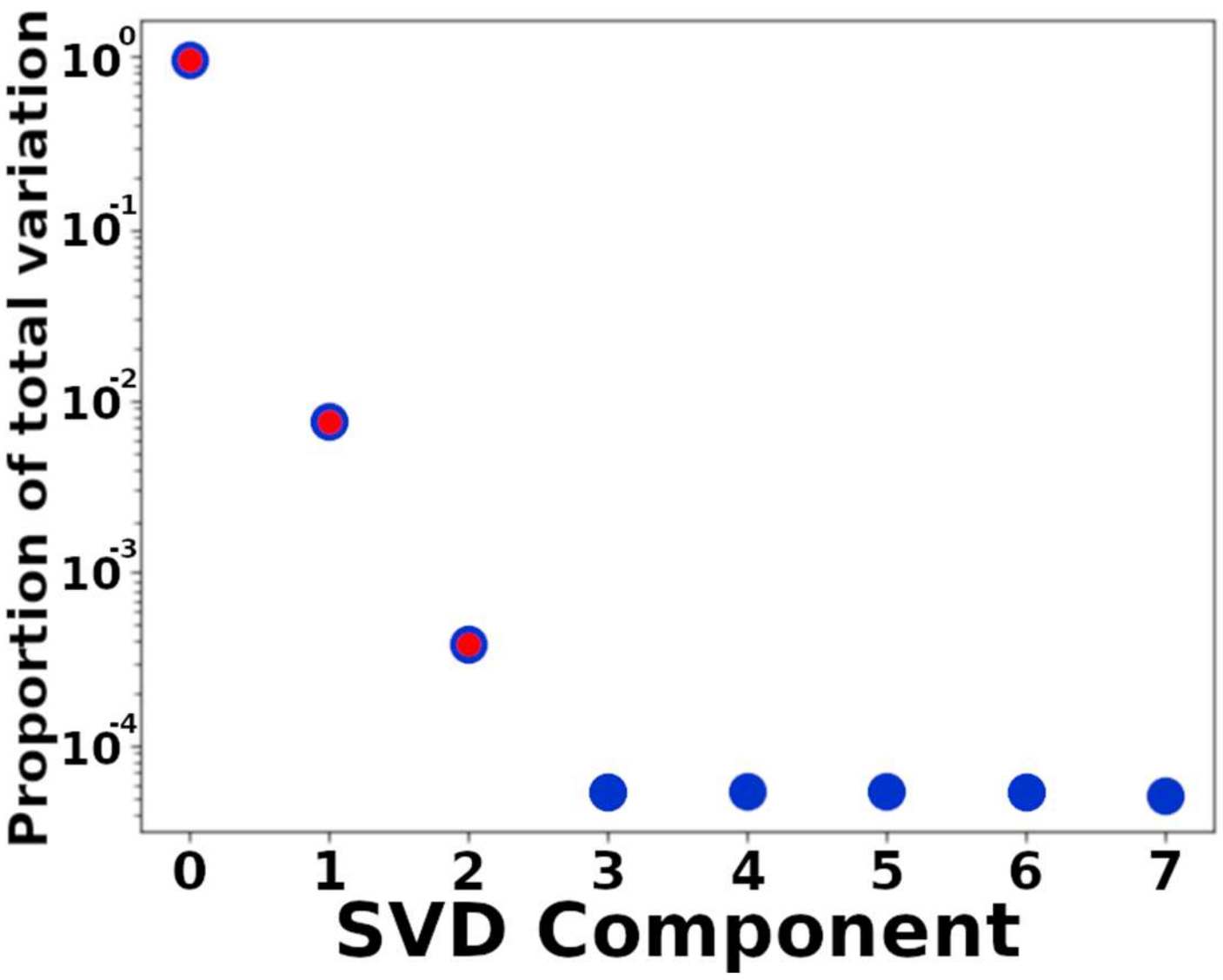

*Figure S2: Scree plot for the masked Singular Value Decomposition, normalized for Poisson noise, of the F-CL dataset. Results show 3 main components in the decomposition.*

Having three meaningful components in this dataset from a chemical point of view (both iron species as well as the background signal from other edges), this makes the F-CL dataset a perfect candidate for the abovementioned BSS method.

## *2. Determination of the Fe-$M_{2,3}$ EELS edges for FeO and $Fe_3O_4$*

We consider that all possible spectra within the F-CL SPIM can be expressed as a combination of a power law background and the normalized, background-extracted Fe-$L_{2,3}$ references gathered from the dataset (here called $L_{FeO}$ and $L_{Fe3O4}$). Similarly, all possible spectra within the F-LL SPIM can be expressed as a combination of a power law background and the spectral fingerprints for FeO and $Fe_3O_4$ ($M_{FeO}$ and $M_{Fe3O4}$), which we have yet to determine. This process is equivalent for background-subtracted SPIMS, with the background-extracted spectra being linear combinations of their respective reference fingerprints.

Both the F-CL and the F-LL SPIMS were subjected to a background removal. In the case of the F-LL SPIM, a dedicated background removal was performed for the Fe-$M_{2,3}$ edge. Thanks to the dataset's high SNR following SVT denoising, we were able to fit a conventional power-law background model on a remarkably narrow pre-edge window spanning 50 to 52 eV.

We consider two points in the background-extracted SPIM, $P_C$ and $P_S$, with their respective Core-Loss EELS spectra, $C_C$ and $C_S$, the first one on the central area of one of the nanoparticles (and therefore on top of both its core and the shell) and another one on the peripheral regions of the nanoparticle (and therefore on top of its shell). In this case, it can be proven that:

$$C_C = a \cdot L_{FeO} + b \cdot L_{Fe_3O_4}$$

$$C_S = c \cdot L_{FeO} + d \cdot L_{Fe_3O_4}$$

We can easily find an expression for $L_{FeO}$ and $L_{Fe3O4}$ as a function of $C_C$ and $C_S$:

$$L_{FeO} = \frac{d}{ad-bc} \cdot C_C + \frac{-b}{ad-bc} \cdot C_s = \alpha \cdot C_C + \beta \cdot C_s$$

$$L_{Fe_3O_4} = \frac{-c}{ad-b} \cdot C_C + \frac{a}{ad-b} \cdot C_s = \gamma \cdot C_C + \delta \cdot C_S$$

The values of $\alpha$, $\beta$, $\gamma$ and $\delta$ were calculated by modelling $L_{FeO}$ and $L_{Fe_3O_4}$ as a sum of $C_C$ and $C_S$, using least square curve fitting. The results of this fit can be seen in Figure S3.

Considering that the ratio between the ELNES features of FeO and $Fe_3O_4$ are equivalent for the Fe-$L_{2,3}$ and Fe-$M_{2,3}$ edges, we can then find the references for these two states in the Fe-$M_{2,3}$ edge by comparing the background-extracted Soft Core-loss EELS spectra taken in the same $P_C$ and $P_S$ points ($S_c$ and $S_S$, respectively):

$$M_{FeO} = \alpha \cdot S_C + \beta \cdot S_S$$

$$M_{Fe_3O_4} = \gamma \cdot S_C + \delta \cdot S_S$$

The full flow chart, as well as the results for the Fe-$M_{2,3}$ edge, can be seen in Figure S3.

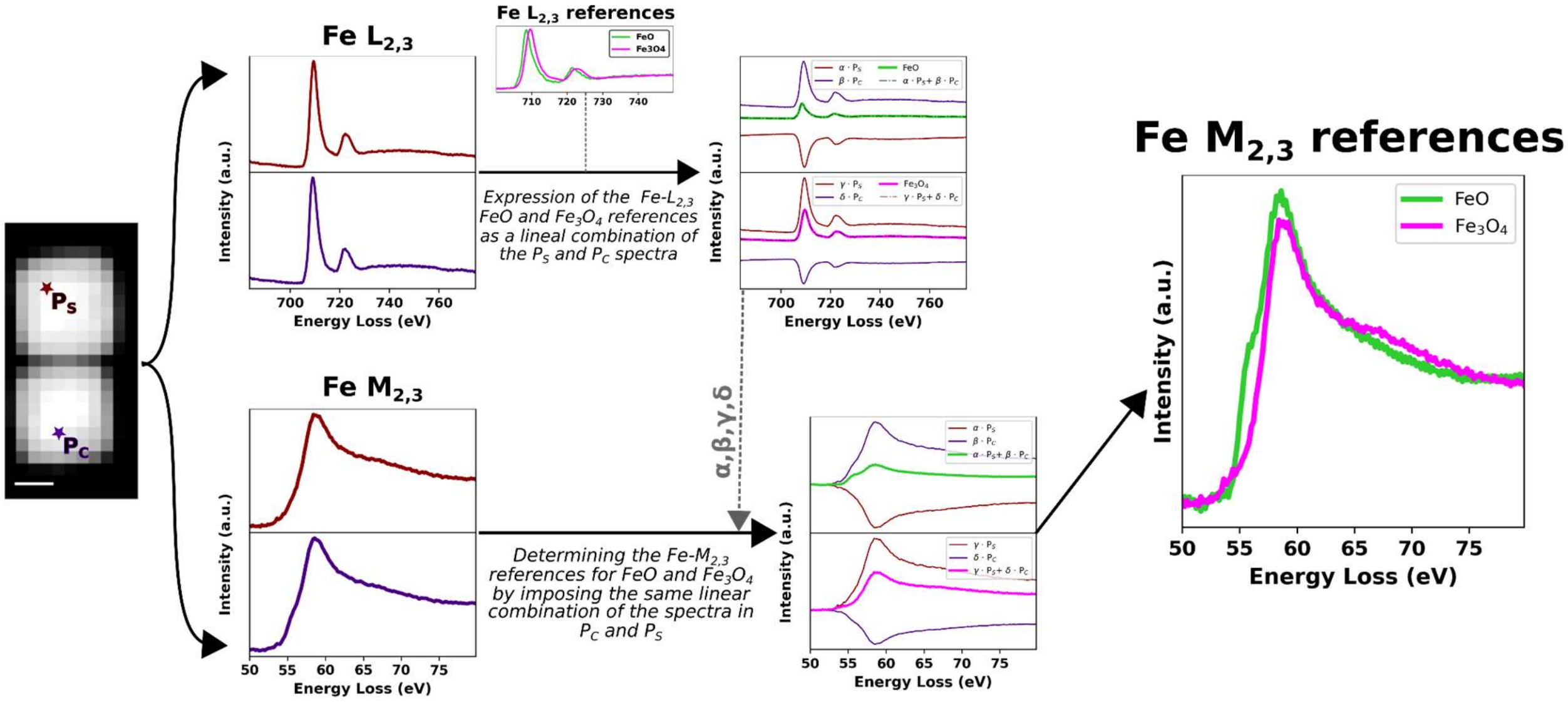

*Figure S3: Flow chart for our custom procedure to determine the FeO and $Fe_3O_4$ Fe-$M_{2,3}$ references, taking their Fe-$L_{2,3}$ references as input.*

### *3. Dose estimation*

The estimation for the dose used two experimental parameters: the number of counts of a pixel in the SPIM where there was only vacuum (or, in the case that was not feasible, on top of the substrate), which we will call $C_T$ and the estimated size of the nanocube being probed.

The probe size was estimated as a combination of the defocus of the microscope itself, $D_M$(estimating a ~1Å defocus), and the geometric defocus caused by the interaction of the beam with the NP, $D_G$. The expression for the probe size would then be:

$$P_s = \pi \cdot D^2$$

With $D$ being the total defocus:

$$D = \sqrt{{D_M}^2 + {D_G}^2}$$

Where $D_G$ is the geometrical defocus:

$$D_G = tan\,(\alpha) \cdot L_{NP}$$

Where $\alpha$ is the convergence angle and $L_{NP}$ the size of the analyzed NP.

Once the probe size has been estimated, the total dose would be a simple quotient, $C_T/P_S$, divided by a factor calculated in the microscope at equivalent conditions, since the number of counts registered is about 30% higher than the actual number of electrons arriving at the detector. As for the dose rate, it would suffice to divide the total dose by the dwell time of the acquisition.

### *4. Initial volume estimations*

In order to have an initial estimation of the FeO/$Fe_3O_4$ volume ratio, a very simple approximation was employed, by taking measurements of the width and height of the shell and the core in the X-Y plane, and approximating their Z measurements as the mean of these two values, assuming a cubic shape.

## *5. DIPm-TV training and network architecture:*

The simulated data reconstructions were performed using 1000 iterations, selecting the reconstruction with the lowest loss. For the experimental data, 2000 iterations were conducted to minimize noise and mitigate overfitting. The network input noise is sampled from a uniform distribution. All reconstructions, model training, and applications were performed on a workstation equipped with an NVIDIA L40 GPU (46 GB of memory).

### Neural network: 3D U-Net

The neural network is a 3D U-Net encoder-decoder with skip connections. Each encoder layer consists of two convolutional layers with a kernel size of 3×3×3 followed by ReLU activations. Max-pooling is applied after each encoding block, and upsampling is performed at each layer of the decoding path. The number of features in the encoding layers progressively increases as [16,32,64,128]. The corresponding number of features in the skip-connections between the encoder and decoder is [4,8,16,32].

### Training details: optimization and convergence

The stability and convergence of the DIPm-TV optimization depend on several hyperparameters, which are described below and summarized in Table 1. The input noise tensor $z$ is initialized with a fixed standard deviation to provide sufficient variability at initialization. The input depth corresponds to the number of channels of the input noise tensor. Network parameters are optimized using the AdamW optimizer [ref], with the learning rate controlling the parameter update step size. In addition, we apply an input-noise regularization strategy by adding small random perturbations to $z$ at each iteration. This acts as an implicit regularizer, reducing overfitting and promoting smoother reconstructions.

The number of iterations determines the number of optimization steps. In the multi-channel configuration, more iterations are typically required to account for the higher complexity of the reconstruction process. Although a fixed number of iterations was used, the final iteration does not necessarily yield the best reconstruction. Therefore, intermediate reconstructions were saved throughout the optimization, and the best checkpoint was manually selected by visual inspection a posteriori based on the clearest structural definition, before noise overfitting appears.

Early stopping is essential when using DIP-based approaches to prevent fitting noise, and to date there is no universally reliable criterion to automatically determine the optimal stopping point. Finally, the TV regularization weight was empirically tuned by exploring different values within a suitable range. The method was implemented using Python, the PyTorch, the Tomosipo module, the ASTRA Toolbox and the corresponding library was released as open-source (https://github.com/CEA-MetroCarac/DL_etomo).

| DIP training hyperparameters | |
|---|---|
| Initial input gaussian noise level | 0.1 |
| Input depth | 32 |
| Learning rate | $10^{-3}$-$10^{-4}$ |
| Noise regularization | 0.01-0.1 |
| Number of iterations | 2000 |
| TV regularization weight | $10^{-8} - 10^{-9}$ |

Table S1: DIP training hyperparameters used to train the network.

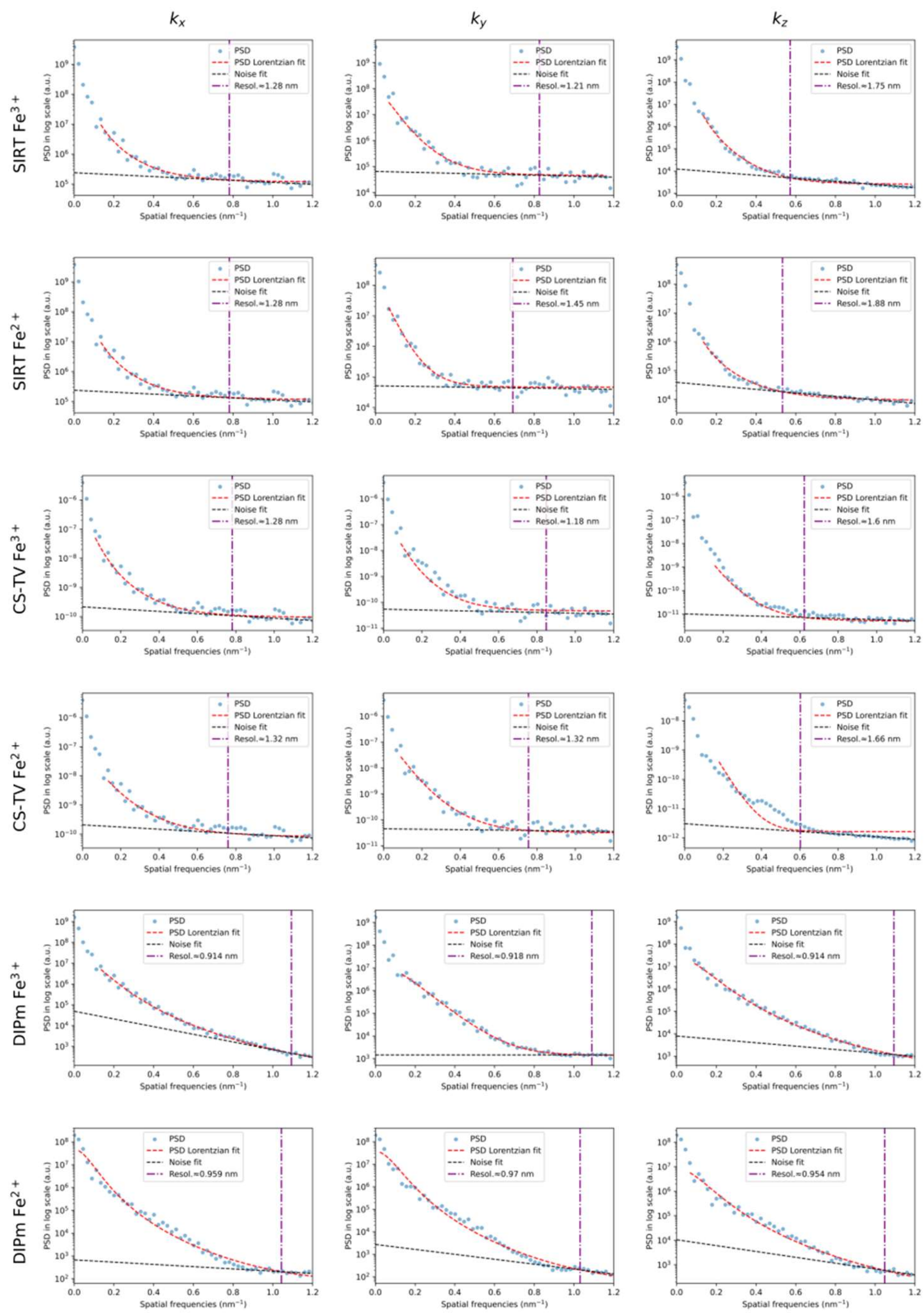

Figure S4. Power spectral density (PSD)-derived directional spatial-frequency cutoffs from 1D PSD profiles computed along $k_x$, $k_y$ and $k_z$ to assess directional resolution in the experimental tomographic reconstructions with SIRT, CS-TV, and DIPm-TV. Each panel shows the PSD profile with a Lorentzian fit (red) and a power law fit to the noise tail (black). The intersection defines the cutoff frequency, from which the corresponding spatial resolution is calculated.

### Power spectral density (PSD)-based directional resolution estimation:

To quantify directional resolution from the reconstructed volumes, we estimated a power spectral density (PSD)-derived spatial-frequency cutoff from 1D PSD profiles computed along $k_x$, $k_y$ and $k_z$. For each axis, we computed the 1D PSD by Fourier transforming the volume and extracting line profiles through the origin along the corresponding frequency axis (followed by averaging to reduce variance). Each 1D PSD was fitted in log-space using a Lorentzian-like roll-off model ($PSD(k)$) plus a constant offset accounting for an approximately white noise floor:

$$PSD(k) = \frac{A}{[1 + (2\pi k\xi)^2]^p} + C \quad (8)$$

where $A$, $\xi$ and $p$ describe the signal decay and $C$ accounts for the high-frequency noise floor. In addition, the noise-dominated tail was modeled by fitting a log-linear function, $log_{10}N(k) = a + b \cdot k$, to the final portion of the PSD tail; yielding $N(k) = 10^{a+b\cdot k}$. The cutoff frequency $k_{cut}$ was defined as the intersection between the fitted PSD model and the estimated noise, i.e. $PSD(k_{cut}) = N(k_{cut})$. This provides an operational resolution limit beyond which the reconstruction becomes noise-dominated and high-frequency details are not reliably recovered (Schwartz et al., 2024). Finally, the effective resolution was computed as $r \approx 1/k_{cut}$ (in nm, with $k$ expressed in $nm^{-1}$).

## *6. DIPm-TV simulation:*

Figure S4 compares SIRT, CS-TV and DIPm-TV reconstructions of the simulated core-shell phantom across different noise levels. At the lowest noise level (Figure S4e-j), the predominant source of artefacts is the combination of the large tilt increment and the mild missing wedge (aligned vertically in these figures). While CS-TV provides a better denoising effect on the reconstructed core and inner shell compared to SIRT, both methods fail to recover the thin outer-shell due to these combined effects. DIPm-TV successfully retrieves all the details in an isotropic manner: the thin outer layer is reliably recovered, both horizontally and vertically.

As the noise level increases, SIRT reconstructions become highly corrupted by noise, while CS-TV produces core and shell structures that are both thicker than the references, with visible staircase artefacts (Figure S4y–z), highlighting the limitations of this approach. In contrast, DIPm-TV reconstructs the core and shell morphology with high fidelity and maintains a smooth appearance even under the noisiest conditions. Most importantly, DIPm-TV reliably resolves the thin outer shell across all noise scenarios, including along the missing wedge direction. This indicates that DIPm-TV can preserve fine structural details under highly sparse-view and low-dose conditions (i.e., high-noise regimes).

This performance is also reflected in Figure S4c-d: DIPm-TV remains more stable and degrades more slowly than CS-TV, which exhibits progressively stronger degradation (orange curve, SNR  12.7 dB). In particular, the SSIM gap between DIPm-TV and CS-TV increases from approximately 0.10 at the lowest noise level to nearly 0.30 at the highest noise level. SIRT is clearly overwhelmed by the combined effects of noise and severe undersampling. A similar trend is observed for NRMSE, although the degradation is less pronounced than for SSIM. Together, these metrics indicate that DIPm-TV better preserves structural features (higher SSIM) and achieves higher quantitative fidelity (lower NRMSE) compared to SIRT and CS-TV.

To complement the qualitative assessment in Figure S4, we further quantified the direction-dependent resolution using a PSD-based directional resolution estimation. In Figure S4ac-ae, we report the PSD-derived resolution along $x$, $y$, and $z$ assuming a pixel size of $0.4\ nm$ (matching the experimental sampling). These results corroborate that DIPm-TV yields higher effective resolution and reduced anisotropy compared with SIRT and CS-TV.

In order to evaluate more precisely DIPm-TV relative to CS-TV for the thin outer shell, Figure S5 shows horizontal intensity profiles of these methodologies for the core signal and three noise scenarios (22.7 dB, 12.7 dB and 8.8 dB), compared against the reference reconstruction. These profiles highlight the strong capability of DIPm-TV across all noise levels, as it accurately matches the thickness of the thin shell layer, and even in the noisiest case the deviation remains small. In addition, the layer remains clearly visible and its shape is preserved. Nevertheless, as noise increases, DIPm-TV exhibits a loss in quantitative fidelity, meaning that the expected intensity of the thin shell is underestimated. Importantly, when the profiles are plotted after denormalization, DIPm-TV preserves the absolute signal levels, indicating that the method remains suitable for quantitative analysis beyond qualitative morphological recovery. In contrast, CS-TV is unable to reconstruct the thin shell. The signal in this region increases only slightly and nearly uniformly over the entire shell region, thereby completely masking the thin outer shell. At the highest SNR (Figure S5a red curve), CS-TV appears to partially “hint” at the thin shell by increasing intensity near the expected location, but still exhibiting clear staircasing.

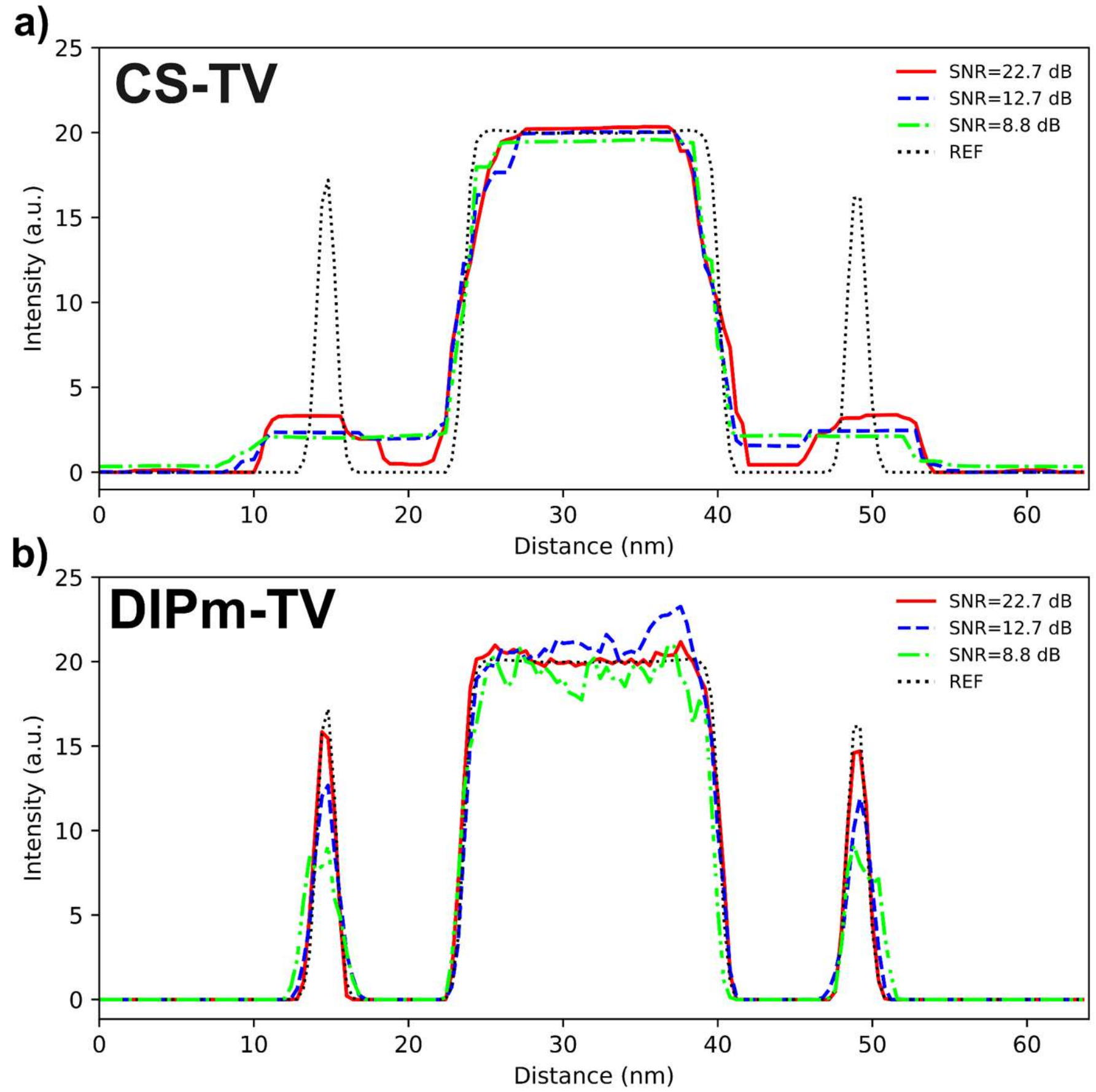

*Figure S5. 1D horizontal line profiles extracted from the central slice of the core channel in Figure S4 of the simulated core–shell phantom, comparing (a) CS-TV and (b) DIPm-TV reconstructions for three noise levels (22.7, 12.7, and 8.8 dB). Colored curves correspond to the profiles at each SNR, while the reference (REF) profile is shown as a black dotted line.*

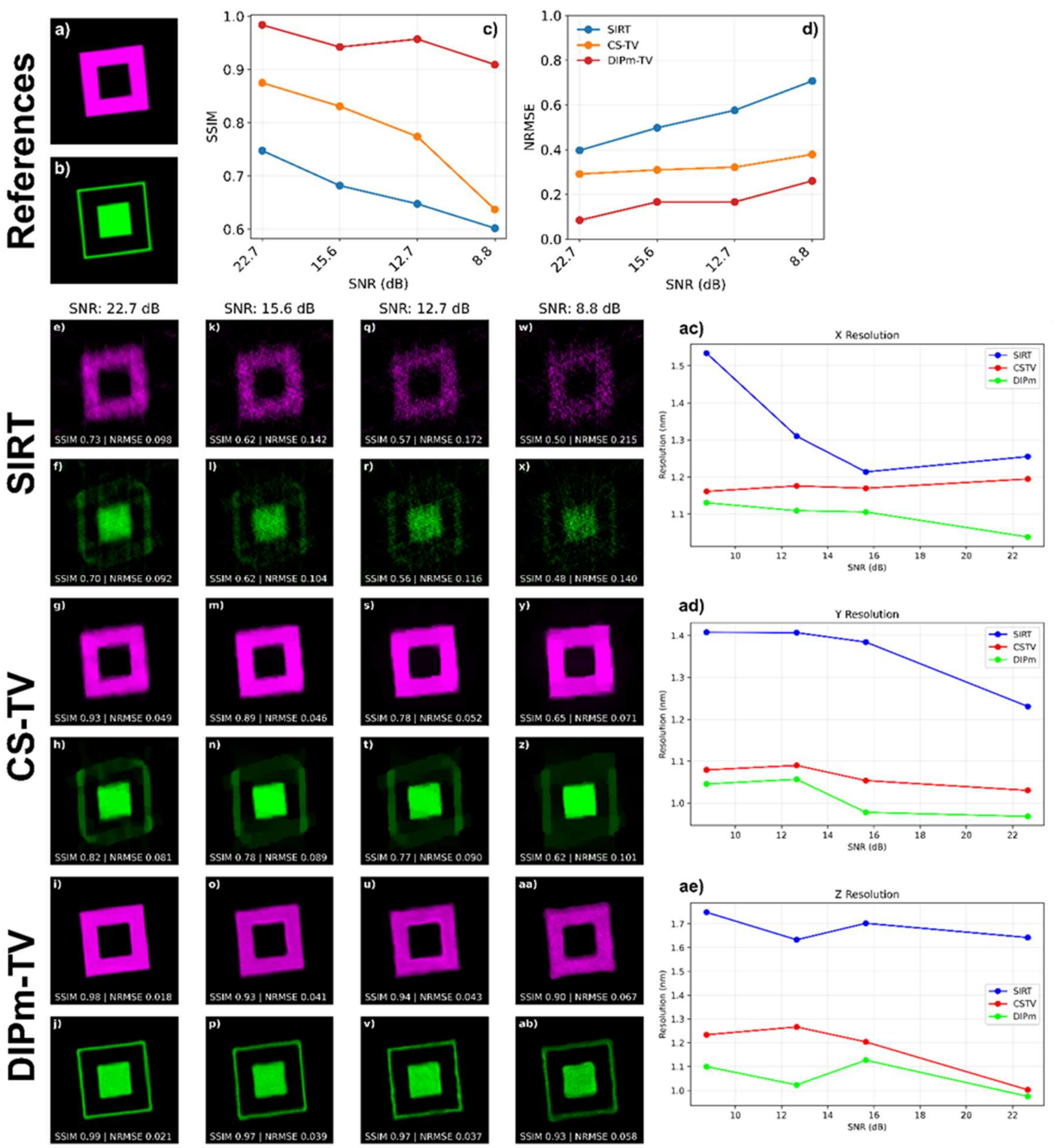

Figure S6 Noise robustness of reconstruction methods for the simulated core-shell phantom, considering the same angular conditions as the EELS tomography experiment [-70°:17.5°:+70°]. (a-b) Reference central cross-section view of the core-shell phantom, reconstructed from SIRT using [-90°:1°:+90°]. (c-d) Quantitative volumetric metrics versus projection SNR: (a) SSIM and (b) NRMSE for SIRT, CS-TV, and DIPm-TV. (e-ab) Central cross-section slices for each SNR level, comparing SIRT (1st-2nd rows), CS-TV (3rd-4th rows) and DIPm-TV (5th-6th rows). (ac-ae) PSD-based directional resolution per method and noise level. SSIM and NRMSE values shown in each view correspond to volumetric measurements.